\documentclass[prx,twocolumn,nofootinbib,citeautoscript,longbibliography,notitlepage,superscriptaddress]{revtex4-1}
\usepackage{csquotes}
\usepackage{graphicx}
\usepackage{dcolumn}
\usepackage{bm}
\usepackage{color}
\usepackage{amsmath}
\usepackage{tabularx}
\usepackage{epstopdf}
\usepackage{booktabs}
\usepackage{nicematrix}
\usepackage{latexsym}
\usepackage{amssymb}
\usepackage{amsmath}
\usepackage{colortbl}
\usepackage{psfrag}
\usepackage{bbm}
\usepackage{bm}
\usepackage{titlesec}
\usepackage{dsfont}
\usepackage{feynmp}
\usepackage{slashed}
\usepackage{multirow}
\usepackage[tight]{subfigure}

\usepackage{comment}

\usepackage[papersize={8.5in,11in}]{geometry}
\usepackage{color}
\definecolor{darkblue}{rgb}{0.,0.,0.4}
\definecolor{darkred}{rgb}{0.5,0.,0.}
\definecolor{BlueViolet}{RGB}{138,43,226}
\definecolor{SkyBlue}{RGB}{30,144,255}
\definecolor{DarkGreen}{RGB}{0,100,0}
\usepackage[pdftex,colorlinks=true,linkcolor=darkblue,citecolor=blue,urlcolor=darkred]{hyperref}

\renewcommand{\epsilon}{\varepsilon}

\allowdisplaybreaks[3] 
\usepackage{CJKutf8}

\usepackage{amsmath}
\usepackage{feynmp}
\usepackage{amssymb}
\usepackage{epsfig}
\usepackage{graphicx}
\usepackage{subfigure}
\usepackage{mathrsfs}
\usepackage{cleveref}
\usepackage{cases}
\usepackage{bbm}
\usepackage{xcolor}
\usepackage{tabularx}
\usepackage{array}
\usepackage{booktabs}
\usepackage{multirow}
\usepackage{lipsum}
\usepackage{float}
\usepackage{lmodern}
\usepackage{textcomp}
\usepackage{makecell}
\usepackage{comment}
\usepackage{threeparttable}
\usepackage{orcidlink}  

\allowdisplaybreaks[4]

\begin{document}
\begin{CJK*}{UTF8}{gbsn}
\title{Massive scalar field perturbations in noncommutative-geometry-inspired Schwarzschild black hole}

\author{Wen-Hao Bian (边文浩)\,\orcidlink{0009-0005-9980-3376}}
\email[]{whbian@smail.nju.edu.cn}
\affiliation{School of Physics, Nanjing University, Nanjing, Jiangsu 210093, China}

\author{Zhu-Fang Cui (崔著钫)\,\orcidlink{0000-0003-3890-0242}}
\email[Contact author: ]{phycui@nju.edu.cn}
\affiliation{School of Physics, Nanjing University, Nanjing, Jiangsu 210093, China}

\date{\today}


\begin{abstract}
In this paper, based on noncommutative-geometry-inspired Schwarzschild black hole, we employ a third-order WKB approximation approach to systematically calculate the quasinormal mode frequencies (QNFs), greybody factors (GFs), and absorption cross section (ACS) under massive scalar field perturbations. The results show that the QNFs satisfy \(\operatorname{Im}(\omega)<0\), confirming the stability of the black hole under perturbations. Furthermore, increasing the noncommutative parameter \(\theta\) reduces the absolute values of both the real and imaginary parts of the frequency, while increasing mass \(\mu\) increases the real part and reduces the imaginary part. The GFs and ACS increase with increasing \(\theta\) and decrease with increasing \(\mu\), indicating opposite modulation effects of these two types of parameters. It is worth emphasizing that the QNFs of the extreme black hole approach the corresponding values of the classical Schwarzschild black hole at angular quantum number $\ell=1$ and large $\mu$, suggesting that, the effects of mass and noncommutative geometry quantum corrections cancel each other out to some extent. It is hoped that these results provide a viable theoretical basis for both the theoretical and experimental aspects of the perturbative dynamics of black hole.
\end{abstract}


\maketitle


\section{Introduction}
In recent years, noncommutative geometry, as an important phenomenological framework for quantum gravity, has attracted widespread attention in the black hole physics. By introducing coordinate noncommutativity, this framework achieves the deconstraint of spacetime at the Planck scale, thereby naturally eliminating the curvature singularity in classical general relativity and offering a potential solution to the black hole information loss paradox~\cite{Hawking2005PRD}. Subsequently, Nicolini et al. constructed Schwarzschild black hole solutions inspired by noncommutative geometry based on Gaussian and Lorentzian matter distributions~\cite{Piero2006PLB,Nicolini2009IJMPA,Ghosh2018CQG,Filho2025PDU,Filho2025AP,Filho2025JCAPKBBH,Filho2025JCAPneutrinos}, which were later generalized to charged, rotating, and higher-dimensional cases, forming a rich family of black hole solutions~\cite{Rizzo2006JHEP,Stefano2007PLB,Euro2009PLB,Modesto2010PRD,Cox2023CQG,Ma2024EPJP}. This framework has found various applications in black hole physics~\cite{Ding2010PRD,Ding2011JHEP,Wei2015JCAP,Sharif2016zEPJC,Batic2019EPJP,Ovgun2020MPLA,Filho2024PDU,Batic2024EPJC,Heidari2025JCAP}.

Quasinormal modes (QNMs) are generated by gravitational waves during the ringdown decay phase of black hole mergers ~\cite{Vishveshwara1970Scattering,Berti2007PRD1}, characterized by a complex natural frequency known as the quasinormal mode frequencies (QNFs). The real and imaginary parts of the QNFs are related to the oscillation and damping of the perturbation, respectively. Since QNMs depend solely on black hole parameters, studying them can help us gain a more precise understanding of black hole~\cite{Fernando1989PRD,Berti2006PRD,Berti2007PRD,Huang:2021qwe}. Consequently, QNMs have long been an attractive research topic.

Since QNFs were introduced by Press, this research has become a central component of black hole perturbation theory~\cite{Press1971AJL}. So far, numerous methods have been employed to study the QNFs, with the most commonly used being the Wentzel-Kramers-Brillouin (WKB) approximation approach~\cite{Schutz1985AJL,Leaver1985AMPS,Iyer1987PRDI, Iyer1987PRDII,Nollert1992PRD, Froman1992PRD, Gundlach1994PRD,Konoplya2003PRD, Pani2013IJMPA,Matyjasek2017PRD}. Regarding four-dimensional noncommutative-geometry-inspired Schwarzschild black hole (NCG-Schwarzschild BH), several studies have explored QNFs using different methods. Liang applied third-order WKB approximation approach and conclude that this black hole is stable for \(3.6M_P<M<19M_P\) with $M_P$ the Planck mass~\cite{Liang2018CPL,Liang2018CPL2}. Batic et al., however, found that third- to sixth-order WKB approximation approach suffers convergence issues near the mass range \(1.91M_P<M<2.3897M_P\), potentially yielding false instability signals~\cite{Batic2019EPJP}; using spectral methods, they rigorously established linear stability in the absence of mass perturbations~\cite{Batic2019EPJP,Batic2024EPJC}. Hu et al. subsequently examined noncommutative charged black hole with Gaussian and Lorentzian matter distributions: for the Gaussian case, only the charge \(q\) significantly increases the QNFs magnitudes, while in the Lorentzian case both \(q\) and the noncommutative parameter \(\theta\) contribute~\cite{Ma2024EPJP}. Furthermore, Ref.~\cite{Karimabadi2025arXiv} compares two nonminimal coupling models and finds that the Ricci scalar coupling and Einstein tensor coupling yield similar QNMs at low frequencies; however, in the high-overtone and strong-coupling regime, they exhibit unique properties. While Fan et al., in the scattering problems, find that $\theta$ and the Einstein tensor coupling constant $\eta$ suppress the greybody factors (GFs) and absorption cross section (ACS), and verifies that the correspondence between GFs and QNMs holds for large angular momentum~\cite{Fan2025arXiv}. In black hole physics, the GFs describes the transmission probability of Hawking radiation through a potential barrier and directly determines the observed evaporation spectrum ~\cite{Sanchez1978PRD,Andersson1995PRD,Kanti2002PRD,Cardoso2006PRL,Konoplya2019PRD}, whereas the ACS characterizes the efficiency of black hole in absorbing incident waves and is a crucial physical quantity in scattering theory~\cite{Mashhoon1973PRD,Fabbri1975PRD,Ould2025PRD}.

Although the aforementioned studies have deepened our understanding of the QNFs of four-dimensional noncommutative black hole, several key gaps remain to be filled. Existing works have not reported how the mass carried by the scalar field specifically modifies the QNFs, GFs, and ACS. As we known that the effective potential of a scalar field with mass tends to \(\mu^2\) at infinity---resulting in an essentially different QNFs compared to the massless case---and considering that noncommutative geometric correction effects may play a significant role in this context, research on this issue holds clear physical significance~\cite{Ohashi2004CQG,Konoplya2005PLB,Zhidenko2006PRD,Konoplya2006PRD,Konoplya2018PRD,Konoplya2018PRDlonglived,Zhang2019PLB,Aragon2021PRD,Lutfuoglu2026arXiv}. Furthermore, whether there exist competitive or synergistic relationships among the mass $\mu$, the noncommutative parameter $\theta$, the angular quantum number $\ell$, and overtone number $n$ remains a topic lacking systematic research.

This work aims to provide a comprehensive and systematic analysis of scalar field perturbations in the NCG-Schwarzschild BH. We consider scalar field with mass and employ the third-order WKB approximation approach to systematically examine, over the parameter range $3.6M_P<M<19M_P$~\cite{Liang2018CPL,Liang2018CPL2}, the evolutionary responses of QNFs, GFs, and ACS under the combined influence of relevant parameters.

The results show that when \(\ell = 1\) and \(\mu\) is large, the QNMs of the extreme black hole with large $\theta$ approaches that of a classical Schwarzschild black hole. Furthermore, we present complete numerical results for the GFs and ACS, elucidating the competing effects of \(\theta\) and \(\mu\) on the transmission probability and absorption efficiency. Through comparison with the classical Schwarzschild black hole, we reveal the unique physical effects of noncommutative geometric quantum corrections, which reveals an increase in \(\theta\) enhances both transmission and absorption, while mass exerts the opposite effect. Furthermore, at the methodological level, we compare the computational results of the third- and sixth-order WKB approximation approach within the safe parameter range of $3.6M_P < M < 19M_P$. We found that even within this range, the sixth-order WKB approximation exhibits convergence issues near extreme parameters, complementing the phenomenon observed by Batic et al.~\cite{Batic2019EPJP}, prompting us to conduct further verification using other methods in the future. These results hold theoretical significance for understanding the perturbative dynamics of black hole under the influence of quantum gravitational effects. In the future, the computational framework presented in this study could be extended to higher-spin massive perturbation fields, and cross-verification could be performed using other higher-precision computational frameworks.

The rest of this paper is organized as follows. Sec.~\ref{secII} reviews how noncommutative geometry has inspired the fundamental properties of Schwarzschild black hole. Sec.~\ref{secIII} derives the perturbation equation for a massive scalar field and reduces them to a Schr\"odinger-type equation with an effective potential. We presents numerical results for the QNFs in Sec.~\ref{secIV}. Within Sec.~\ref{secV}, we examine the GFs, and Sec.~\ref{secVI} analyzes the ACS. Finally, we summarizes the main conclusions and outlines future research directions in Sec.~\ref{secVII}.

\section{Noncommutative-geometry-inspired Schwarzschild black hole}\label{secII}
Near the Planck scale, the classical description of spacetime is expected to be superseded by quantum gravitational effects. Noncommutative geometry serves as an effective model of quantum gravity phenomenology, positing that spacetime coordinates are fundamentally noncommutative, satisfying $[\hat{x}^\mu,\hat{x}^\nu]=i\theta^{\mu\nu}$, where $\theta^{\mu\nu}$ is an antisymmetric tensor characterizing the degree of spacetime “fuzziness”~\cite{Smailagic2003JPA,Piero2006PLB,Nicolini2009IJMPA,Luo:2014iha,Panotopoulos2020EPJP}. Within the framework of noncommutative geometry, a point-like mass source is replaced by an isotropic Gaussian matter distribution~\cite{Piero2006PLB,Panotopoulos2020EPJP}
\begin{eqnarray}
\rho_\theta(r)=\frac{M}{(4\pi\theta)^{\frac{3}{2}}}e^{-\frac{r^2}{4\theta}},
\end{eqnarray}
which implies that, due to quantum fuzziness of spacetime, the matter energy density is no longer localized at a point but is smeared over a region of characteristic scale $\theta$. This treatment is consistent with the convolution structure in position space found in noncommutative quantum field theory. Substituting the above matter distribution into the Einstein field equation and assuming static spherical symmetry yields the NCG-Schwarzschild metric~\cite{Panotopoulos2020EPJP,Piero2006PLB,Nicolini2009IJMPA,Liang2018CPL,Liang2018CPL2,Batic2024EPJC,Das2019PRD}
\begin{equation}
ds^2=-f(r)dt^2+\frac{1}{f(r)}dr^2+r^2(d\vartheta^2+\sin^2\vartheta d\varphi^2),
\end{equation}
with the metric function,
\begin{equation}
f(r)=1-\frac{4M}{r\sqrt{\pi}}\gamma\left(\frac{3}{2},\frac{r^2}{4\theta}\right),\label{eq:metric}
\end{equation}
where $\gamma\left(\frac{3}{2},\frac{r^2}{4\theta}\right)$ is the lower incomplete Gamma function,
\begin{eqnarray}
\gamma\left(\frac{3}{2},\frac{r^2}{4\theta}\right)
=
\int_0^{\frac{r^2}{4\theta}}\sqrt{t}~e^{-t}dt.
\end{eqnarray}
In the near-horizon region $r\ll \sqrt{\theta}$ , the upper limit of the integral is small, and expanding gives $\gamma(3/2,x)\sim(2/3) x^{3/2}$.  Substituting into the metric function Eq.~(\ref{eq:metric}) simplifies to $f(r)\sim 1-\frac{M r^2}{3\sqrt{\pi}\theta^{3/2}}$. Notably, as \( r \to 0 \), \( f(r) \to 1 \),  in contrast to the divergent behavior \( f(r) \to -\infty \) of the classical Schwarzschild black hole. This indicates that noncommutative geometry effectively smears the curvature singularity, yielding a finite value of the metric function at the origin, which is one of the most important features of noncommutative black holes. In the far-field region $r\gg \sqrt{\theta}$, $\gamma(3/2,x)\to \Gamma(3/2)=\sqrt{\pi}/2$，and the metric asymptotically approaches the standard Schwarzschild form $f(r)=1-2 M/r$. Thus, noncommutative corrections are significant only on quantum scales near the horizon, while in the far-field region the black hole behaves indistinguishably from the classical Schwarzschild black hole, satisfying asymptotic flatness.

\begin{figure}
\centering
\includegraphics[width=3in]{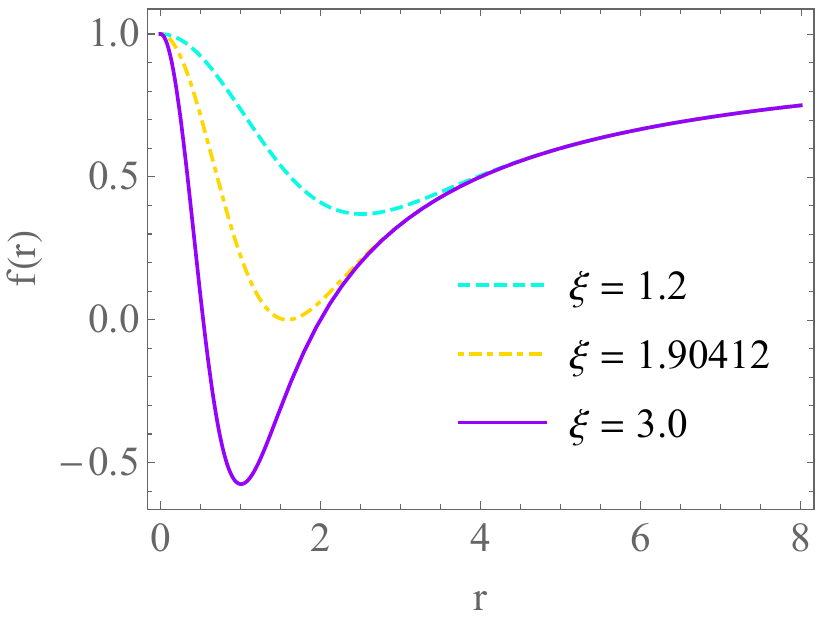}
\caption{(Color online) The metric function \(f(r)\) versus \(r\), for different values of \(\xi \equiv M / \sqrt{\theta}\).}
\label{fig:f_vs_r}
\end{figure}

The horizon of the NCG-Schwarzschild BH is determined by the equation \( f(r_H) = 0 \). Since \( \gamma(3/2, r^2/(4\theta)) \) is monotonically increasing and bounded, the solution to this equation exhibits critical behavior. Defining the dimensionless parameter $\xi\equiv M/\sqrt{\theta}$, the horizon structure can be classified into three cases, as shown in Fig.~\ref{fig:f_vs_r}.

\begin{enumerate}
\item Two-horizon case (\( \xi > \xi_c \approx 1.90412 \)): There exist two horizons, including an outer horizon, which defines the region from which nothing can escape, and the inner horizon arises from the modification of the metric behavior due to quantum corrections and is a common feature of many quantum black hole models.
\item Extreme black hole case (\( \xi = \xi_c \)): One degenerate horizon. In this case, the black hole possesses a minimum non-zero horizon radius \( r_H^{\text{min}} \approx 3.0\sqrt{\theta} \) , so that the surface gravity vanishes, implying zero Hawking temperature.
\item No-horizon case (\( \xi < \xi_c \)): The matter distribution is sufficiently diffuse that no event horizon forms, corresponding to a singularity-free quantum gravitational object.
\end{enumerate}
This implies that, within the noncommutative geometry framework, only black hole with mass above the Planck scale can form classical horizons. This result is qualitatively consistent with expectations from the generalized uncertainty principle (GUP)~\cite{Piero2006PLB,Nicolini2009IJMPA,Kempf1994PRD,Scardigli1999PLB}. Therefore, for a NCG-Schwarzschild BH possessing a horizon, the noncommutative parameter $\theta$ satisfies~\cite{Liang2018CPL}
\begin{eqnarray}
0 < \theta \leq \left( \frac{M}{1.90412} \right)^2.\label{eq:theta_range}
\end{eqnarray}
We naturally set $\sqrt{\theta}=\ell_P$, where the Planck length $\ell_P=\sqrt{\hbar G/c^3}\sim 10^{-35}$m, which correspondingly requires the black hole mass $M>3.6 M_P$~\cite{Batic2019EPJP}. This means that smaller black hole masses allow smaller upper bounds on $\theta$; conversely, for a fixed $\theta$, there exists a lower bound on the black hole mass below which no horizon can form. Following the parameter range of Liang et al.~\cite{Liang2018CPL}, we set $M = 1$ to make the relevant parameters dimensionless, and consequently the noncommutative parameter varies in the range $0<\theta\le 0.2758$. In particular, when $\theta\to 0$, the metric reduces to the standard Schwarzschild metric $f(r)=1-2M/r$, while $\theta\to 0.2758$ corresponds to the extreme black hole. Thus, the NCG-Schwarzschild metric provides us with a well-suited platform to study the QNMs of both the classical Schwarzschild black hole and the quantum-corrected extreme black hole.

\section{Perturbation equation for massive scalar field}\label{secIII}
Consider a scalar field $\Phi$ with mass $\mu$ propagating in the curved background spacetime given by Eq.~(\ref{eq:metric}). Its dynamics is governed by the massive Klein-Gordon equation,
\begin{eqnarray}
\frac{1}{\sqrt{-g}} \partial_\mu \left( \sqrt{-g} g^{\mu\nu} \partial_\nu \Phi \right) - \mu^2 \Phi = 0.\label{eq:KG}
\end{eqnarray}
For the NCG-Schwarzschild metric Eq.~(\ref{eq:metric}), the determinant of the metric can be directly computed as $\sqrt{-g} = r^2 \sin\vartheta$. The corresponding inverse metric components are
\begin{eqnarray}
g^{tt} = -\frac{1}{f(r)},g^{rr} = f(r),g^{\vartheta\vartheta} = \frac{1}{r^2},g^{\varphi\varphi} = \frac{1}{r^2 \sin^2\vartheta}.
\end{eqnarray}
Substituting these expressions into the Klein-Gordon equation Eq.~(\ref{eq:KG}), and noting that the metric function depends only on the radial coordinate \(r\), we obtain
\begin{widetext}
\begin{align}
-\frac{1}{f}\frac{\partial^2\Phi}{\partial t^2} + \frac{1}{r^2}\frac{\partial}{\partial r}\left( r^2 f \frac{\partial\Phi}{\partial r} \right) + \frac{1}{r^2\sin\vartheta}\frac{\partial}{\partial\vartheta}\left( \sin\vartheta \frac{\partial\Phi}{\partial\vartheta} \right) + \frac{1}{r^2\sin^2\vartheta}\frac{\partial^2\Phi}{\partial\varphi^2} - \mu^2\Phi = 0.\label{eq:PDE}
\end{align}
\end{widetext}
Owing to the spherical symmetry of the background spacetime, the scalar field can be expanded in terms of spherical harmonics~\cite{Bagrov1990CQG,Konoplya2018PRD}
\begin{eqnarray}
\Phi(t, r, \vartheta, \varphi) = \sum_{\ell, m} \frac{\psi_{\ell m}(t, r)}{r} Y_{\ell m}(\vartheta, \varphi),\label{eq:Phi}
\end{eqnarray}
where \(Y_{\ell m}(\vartheta, \varphi)\) are the spherical harmonics satisfying $\square_{(\vartheta,\varphi)} Y_{\ell m}=-\ell(\ell+1)Y_{\ell m}$. Substituting the expansion Eq.~(\ref{eq:Phi}) into Eq.~(\ref{eq:PDE}) and performing systematic algebraic manipulations to eliminate the common spherical harmonic factors, we obtain the partial differential equation for the radial wavefunction
\begin{eqnarray}
\ddot{\psi}- f^2 \psi'' - f f'\psi' + \left( \frac{f f'}{r} + \frac{fL}{r^2} + f \mu^2 \right) \psi = 0,
\end{eqnarray}
where $L\equiv-\ell(\ell+1)$ and \(f' \equiv df/dr\). To cast this into a standard form more convenient for wave analysis, we introduce the tortoise coordinate $r_*$, defined by $\frac{dr_*}{dr} = \frac{1}{f(r)}$. In terms of the tortoise coordinate, the wave equation simplifies to
\begin{eqnarray}
\frac{\partial^2 \psi}{\partial t^2} - \frac{\partial^2 \psi}{\partial r_*^2} + V(r) \psi = 0,\label{eq:wave_func}
\end{eqnarray}
with the effective potential \(V(r)\) given by
\begin{eqnarray}
V(r) = f(r) \left[ \frac{\ell(\ell+1)}{r^2} + \frac{1}{r} \frac{df(r)}{dr} + \mu^2 \right].\label{eq:potential}
\end{eqnarray}
This is the standard form of a $(1+1)$-dimensional wave equation, describing the propagation of scalar field perturbations in the tortoise coordinate, where the effective potential $V(r)$ encodes the geometric information of the background spacetime as well as the intrinsic properties of the scalar field.

\begin{figure}
\centering
\subfigure[~$\theta=0.2,~\mu=0$]
{\includegraphics[width=2.4in]{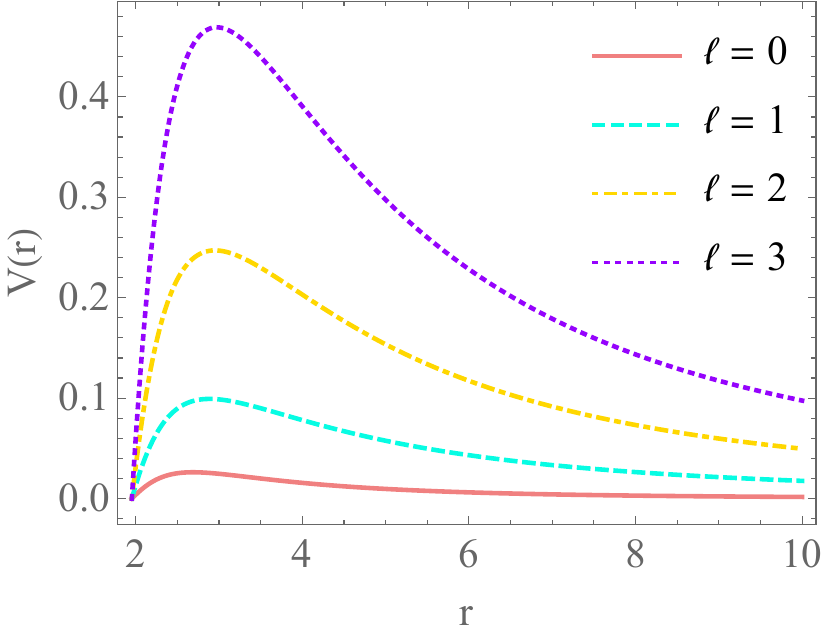}}
\subfigure[~$\ell=0,~\mu=0$]{\includegraphics[width=2.5in]{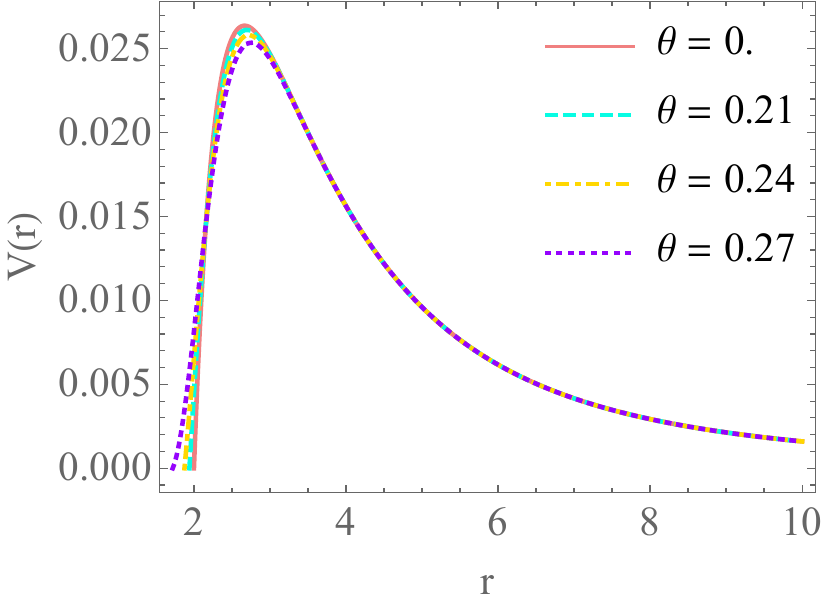}}
\subfigure[~$\ell=0,~\theta=0.2$]{\includegraphics[width=2.5in]{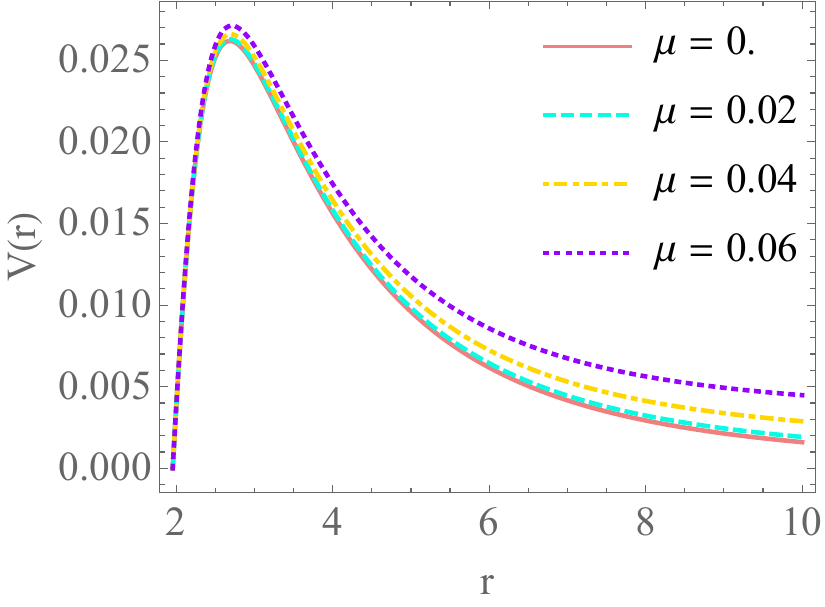}}
\caption{(Color online) The effective potential \(V(r)\) versus \(r\) with different (a) angular quantum number \(\ell\), (b) nonaommutative parameter \(\theta\), and (c) scalar mass \(\mu\).}
\label{fig:V_vs_rstar_without_time}
\end{figure}

The asymptotic behavior of the effective potential $V(r)$ determines the boundary conditions for the perturbation equation. Near the horizon \(r \to r_H\), since \(f(r) \sim (r - r_H) f'(r_H) \to 0\), the effective potential vanishes. At spatial infinity \(r \to \infty\), \(f(r) \to 1\), and the effective potential asymptotically approaches the constant \(\mu^2\). Therefore, the effective potential tends to a positive constant at infinity, which is a key feature distinguishing massive scalar field perturbations from the massless case. To intuitively illustrate the behavior of the effective potential, we examine its dependence on relevant parameters in Fig.~\ref{fig:V_vs_rstar_without_time}. Fig.~\ref{fig:V_vs_rstar_without_time}(a) shows the effective potential for fixed $\theta$ and $\mu$ while varying the angular quantum number $\ell$, revealing that the potential barrier height increases with $\ell$. Furthermore, the noncommutative parameter $\theta$ quantitatively modifies the evolution of the peak of the effective potential, as shown in Fig.~\ref{fig:V_vs_rstar_without_time}(b). For fixed $\ell$ and $\mu$, increasing $\theta$ lowers the potential barrier height. Fig.~\ref{fig:V_vs_rstar_without_time}(c) demonstrates that a larger scalar field mass $\mu$ results in a higher potential barrier, and in the far-horizon region the effective potential approaches the constant $\mu^2$.

To investigate the QNMs for massive scalar field perturbations, we consider a single-frequency mode solution \(\psi(t, r) = e^{-i\omega t} \psi(r)\).  Substituting this into Eq.~(\ref{eq:wave_func}) yields the time-independent Schr\"odinger-type equation,
\begin{eqnarray}
\frac{d^2 \psi}{dr_*^2} + \left[ \omega^2 - V(r) \right] \psi = 0. \label{eq:Schordinger}
\end{eqnarray}
The QNMs are defined by solutions satisfying specific boundary conditions that reflect the physical essence of the black hole perturbation problem. At the horizon, since the event horizon acts as a one-way membrane, waves must be purely ingoing, i.e.,
\begin{eqnarray}
\psi \sim e^{-i\omega r_*}, \quad r_* \to -\infty.
\end{eqnarray}
At spatial infinity, in the asymptotic region, for a massive scalar field the dispersion relation is \(\omega^2 = k^2 + \mu^2\), where \(k\) is the wavenumber. Physically, waves are required to be outgoing (escaping to infinity), namely
\begin{eqnarray}
\psi \sim e^{+i\sqrt{\omega^2 - \mu^2} \, r_*}, \quad r_* \to +\infty,\label{eq:boundry_condition2}
\end{eqnarray}
which means if $\mathrm{Im}(\omega)\sim0$ and $\mathrm{Re}(\omega)<\mu$, no energy transition to infinity~\cite{Konoplya2011RMP}.  These two boundary conditions together constitute an eigenvalue problem: only for certain specific complex values of the frequency $\omega$ do nontrivial solutions that satisfy both boundary conditions simultaneously exist.

\section{Quasinormal modes in the WKB approximation approach}\label{secIV}
Based on the perturbation equations and boundary conditions derived in the previous section, we examine the QNFs of the NCG-Schwarzschild BH under massive scalar field perturbations using the WKB approximation method. Rewriting the Schrödinger-type equation Eq.~(\ref{eq:Schordinger}) as
\begin{eqnarray}
\frac{d^2\psi}{dr_*^2}+Q(r_*)\psi=0,
\end{eqnarray}
where $Q(r_*)\equiv \omega^2-V(r_*)$. Within the WKB framework, the QNFs are determined by the following condition, expanding $Q$ around its maximum $r_*^{(0)}$ yields~\cite{Iyer1987PRDI,Iyer1987PRDII,Konoplya2011RMP},
\begin{eqnarray}
\frac{iQ_0}{\sqrt{2 Q_0''}}-\sum_{i=2}^N\Lambda_i=n+\frac{1}{2},
\end{eqnarray}
where $N$ is the order number, $Q_0$ and $Q_0''$ denote the value and the second derivative of $Q(r_*)$ at its maximum, respectively, $n$ is the overtone number, and $\Lambda_i$ are correction terms that depend on $Q$ and its higher-order derivatives evaluated at the extremum. The third- and higher-order WKB approximation approach is discussed in detail in Refs.~\cite{Iyer1987PRDI,Matyjasek2017PRD,Konoplya2003PRD}.

To ensure the validity of the WKB approximation approach, it is typically required that $\ell>n$~\cite{Liang2018CPL}. In \ref{compare}, we compare the results obtained using the third-order and pure sixth-order WKB approximation approach (see Fig.~\ref{fig:omega_vs_theta_varying_order}). It is observed that when $\theta$ is large (close to the extremal black hole), the convergence of the sixth-order WKB is significantly worse than that of the third-order. This observation is consistent with the findings of Batic et al., who reported that higher-order WKB approximation approach may suffer from convergence issues in extreme regimes and can even yield spurious instability signals~\cite{Batic2024EPJC}. Therefore, all subsequent calculations in this paper are performed using the third-order WKB approximation approach, and future work will employ spectral methods~\cite{Batic2024EPJC} or Pad\'e-improved WKB techniques~\cite{Matyjasek2017PRD,Lutfiuoglu2025EPJC,Konoplya2019CQG} to further validate the results in the extremal regime. 

\begin{figure*}
\centering
\subfigure[~$\ell=1,n=0$]
{\includegraphics[width=1.7in]{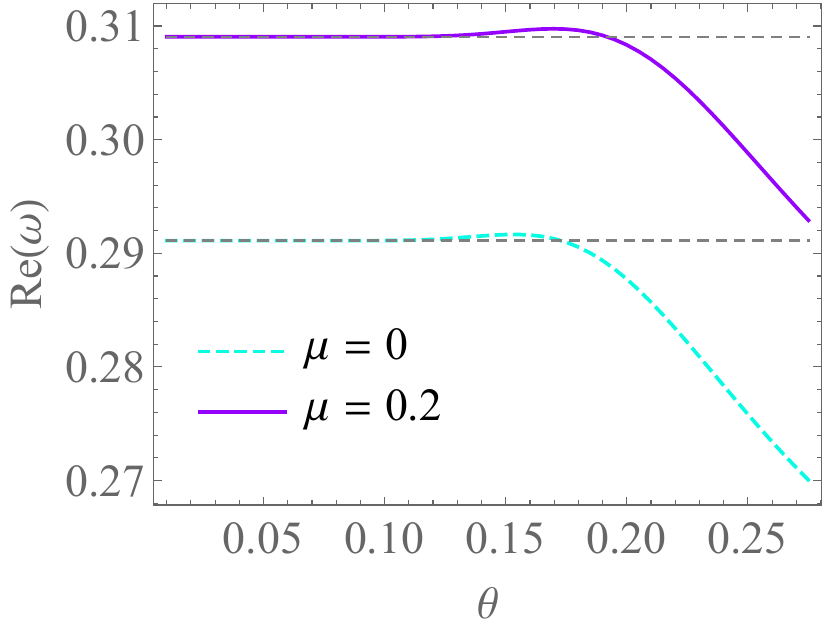}}
\subfigure[~$\ell=1,n=0$]
{\includegraphics[width=1.7in]{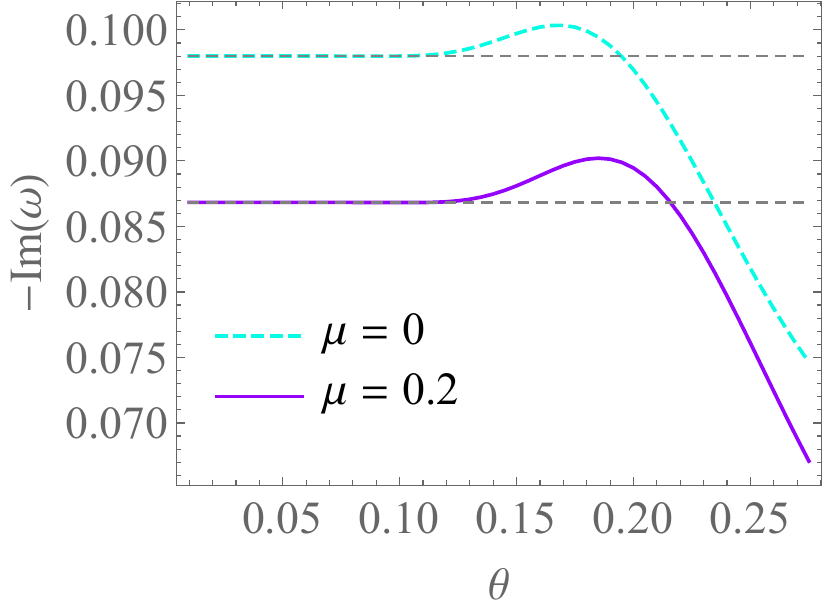}}
\subfigure[~$\ell=2,n=0$]
{\includegraphics[width=1.7in]{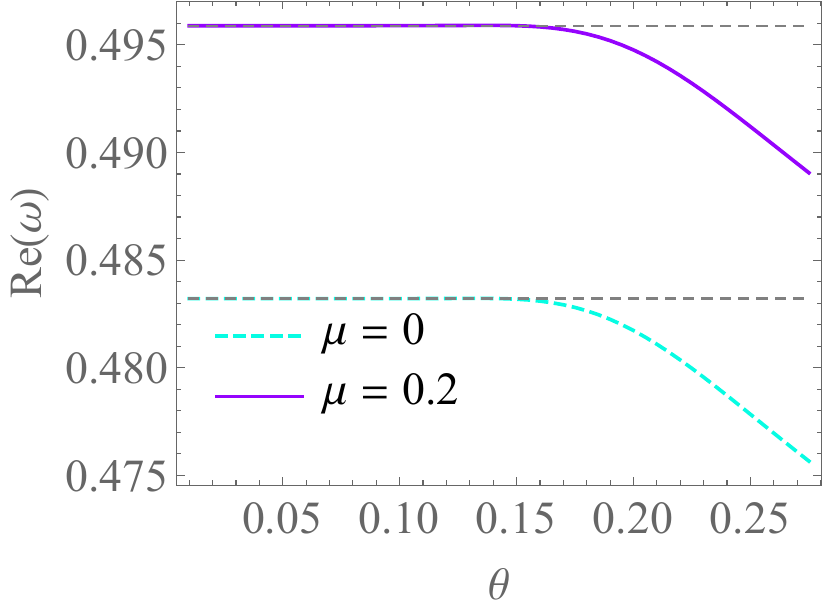}}
\subfigure[~$\ell=2,n=0$]
{\includegraphics[width=1.7in]{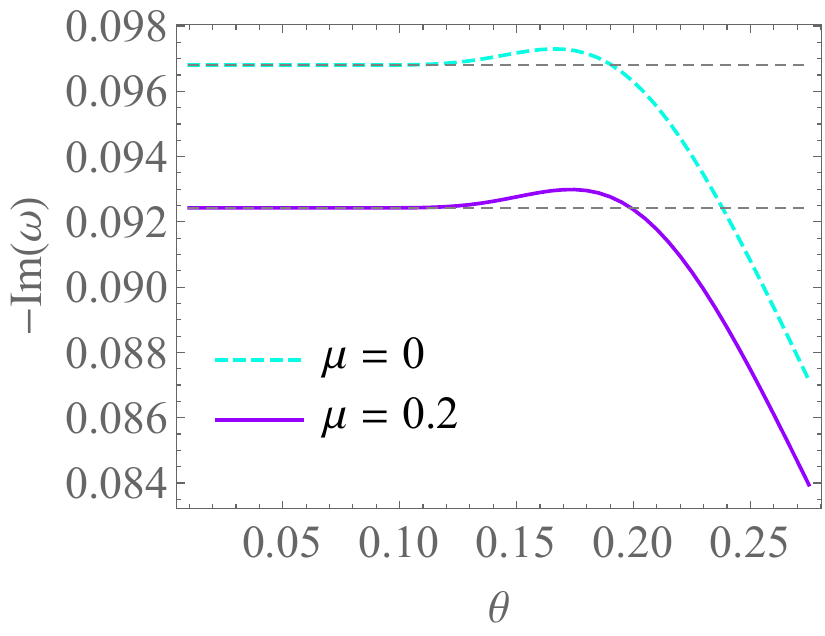}}\\
\subfigure[~$\ell=2,n=1$]
{\includegraphics[width=1.7in]{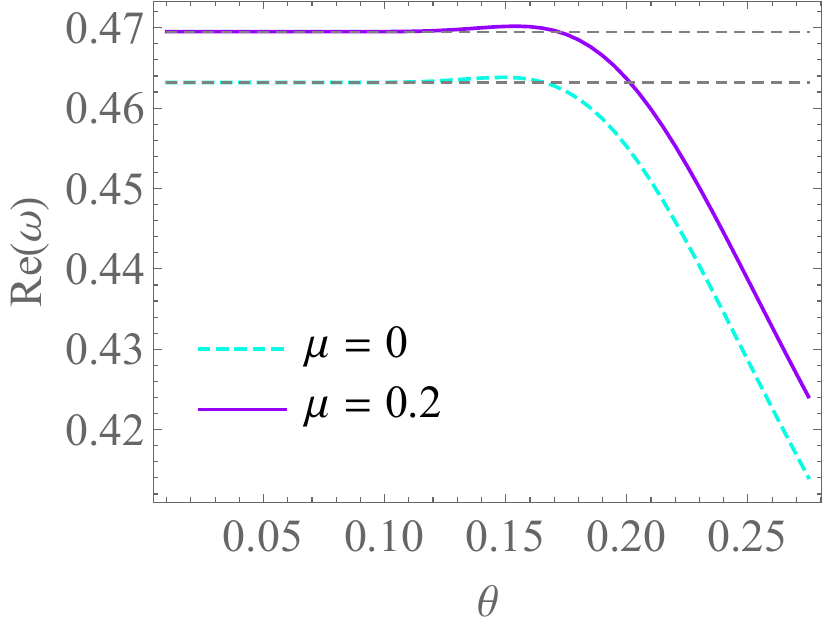}}
\subfigure[~$\ell=2,n=1$]
{\includegraphics[width=1.7in]{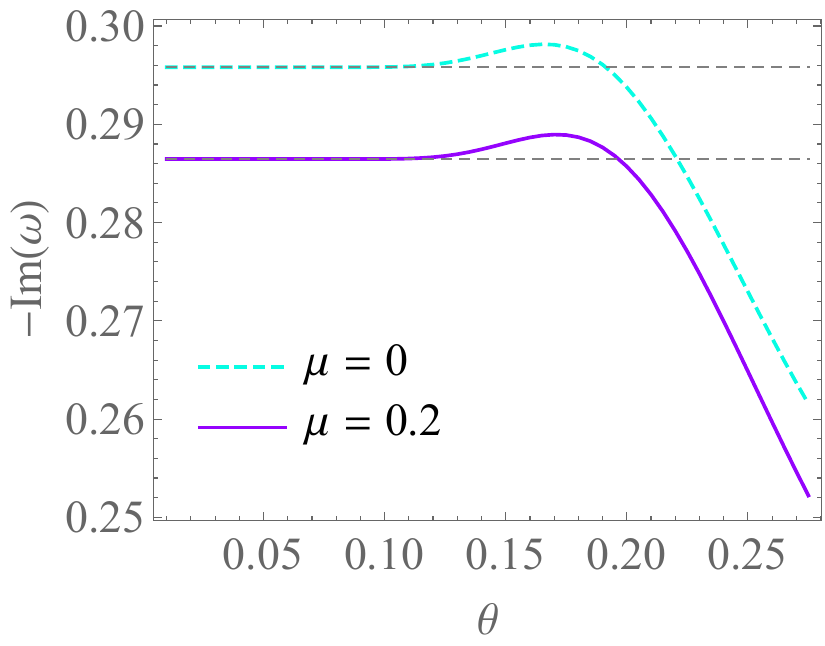}}
\subfigure[~$\ell=3,n=0$]
{\includegraphics[width=1.7in]{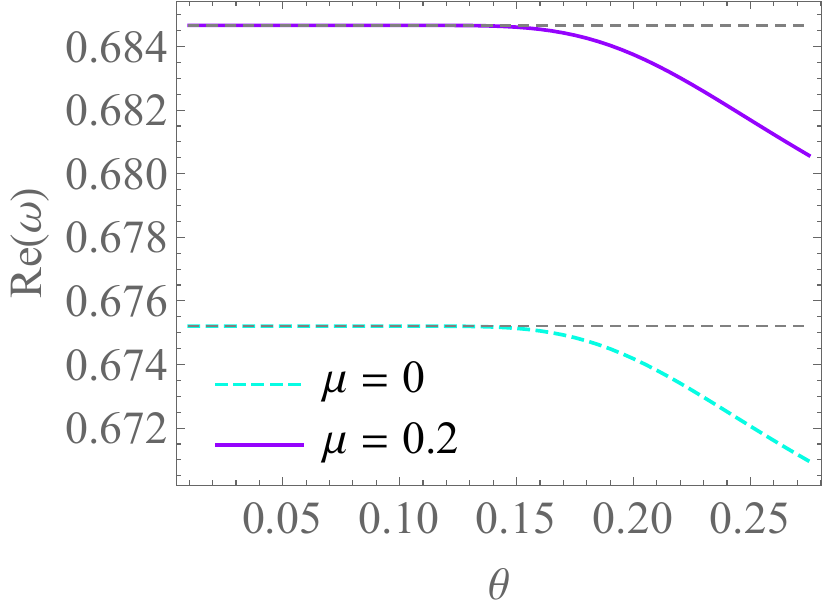}}
\subfigure[~$\ell=3,n=0$]
{\includegraphics[width=1.7in]{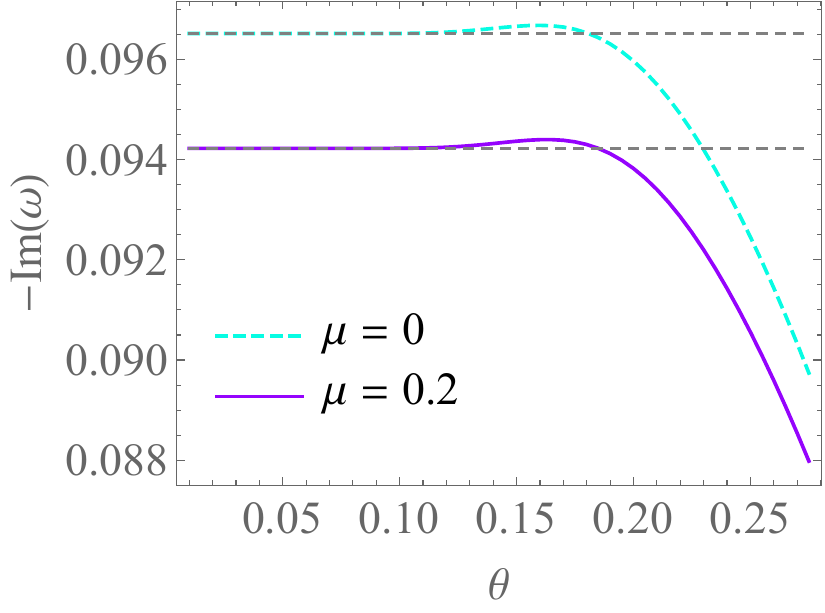}}\\
\subfigure[~$\ell=3,n=1$]
{\includegraphics[width=1.7in]{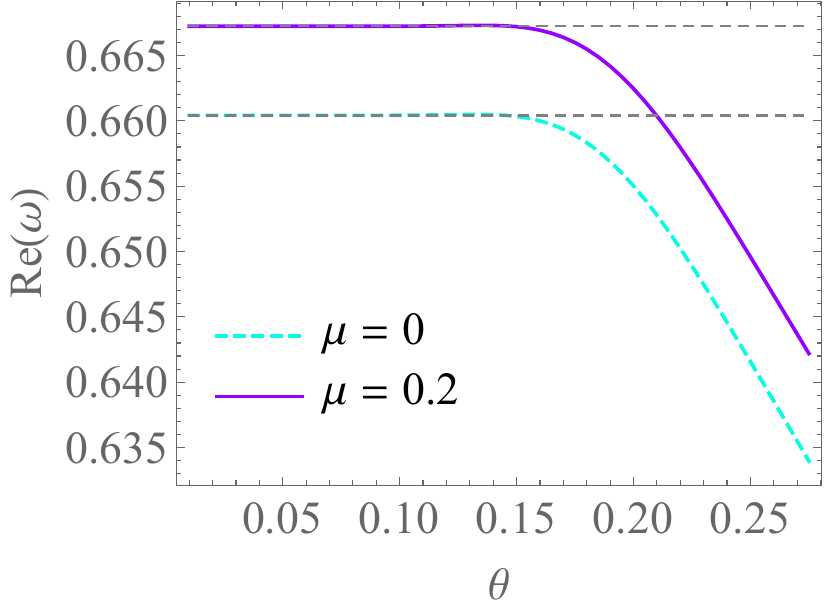}}
\subfigure[~$\ell=3,n=1$]
{\includegraphics[width=1.7in]{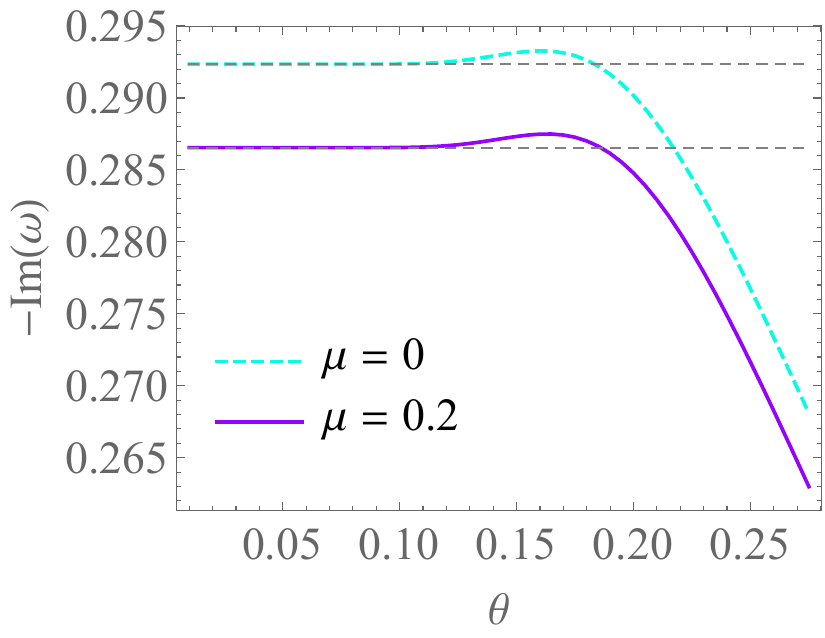}}
\subfigure[~$\ell=3,n=2$]
{\includegraphics[width=1.7in]{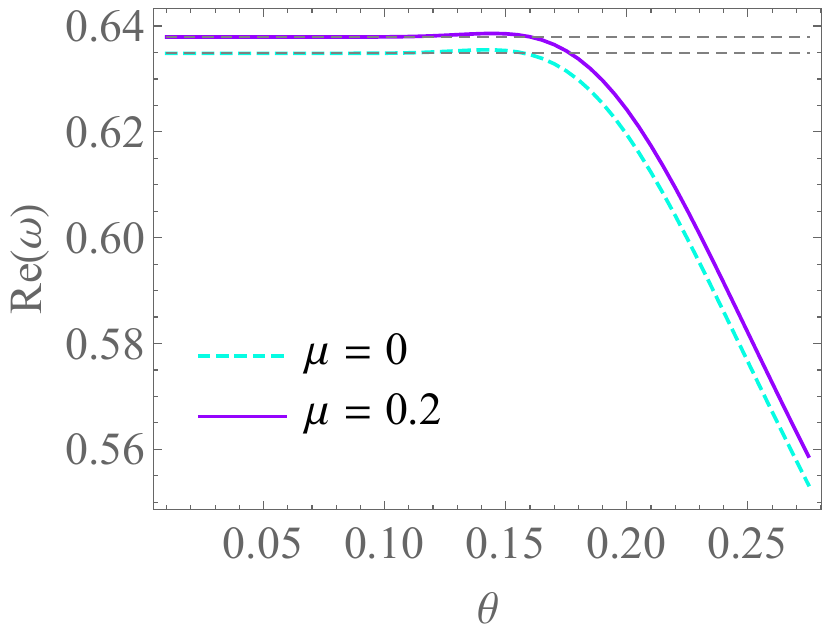}}
\subfigure[$\ell=3,n=2$]
{\includegraphics[width=1.7in]{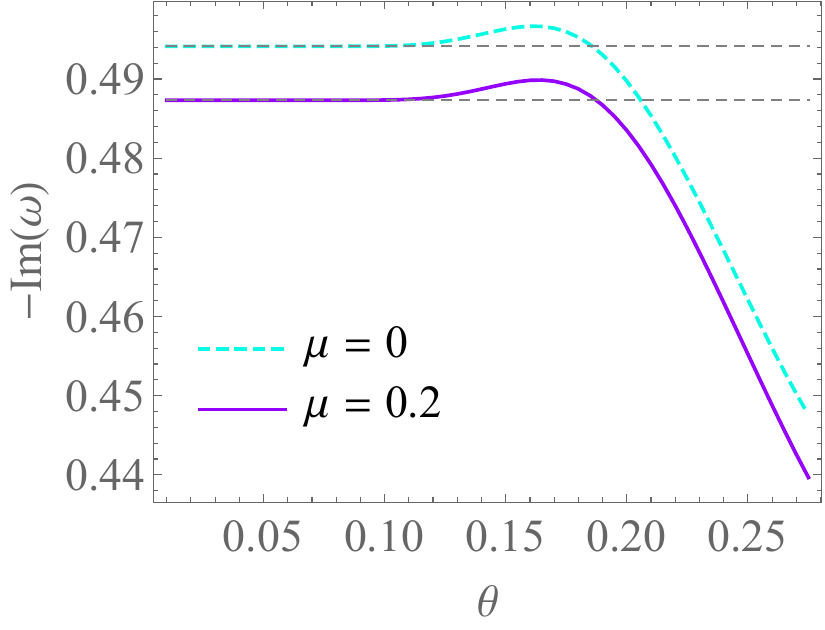}}
\caption{(Color online) The evolution of the real part $\mathrm{Re}(\omega)$ and the imaginary part $-\mathrm{Im}(\omega)$ of QNFs as a function of the noncommutative parameter $\theta$ for representative scalar mass.}
\label{fig:ReIm_vs_theta_without_time}
\end{figure*}

First, in Fig.~\ref{fig:ReIm_vs_theta_without_time}, we present the evolution of QNFs as a function of $\theta$ under various relevant parameters. The results indicate that the findings with mass are qualitatively consistent with those without mass~\cite{Liang2018CPL}. Specifically, all QNFs have negative imaginary parts $\mathrm{Im}<0$,  indicating that the NCG-Schwarzschild BH is linearly stable under massive scalar field perturbations.  The perturbation amplitude decays exponentially with time, with the decay lifetime given by $\tau=1/|\mathrm{Im}(\omega)|$. Furthermore, the noncommutative parameter $\theta$ significantly affects the QNFs. In the two-horizon region with small $\theta$, the deviations from the classical Schwarzschild black hole can be either positive or negative, depending on $\ell$ and $n$. Near the extremal black hole with large $\theta$, both the real part and the absolute value of the imaginary part are always smaller than their classical Schwarzschild counterparts, indicating that scalar field perturbations oscillate and decay more slowly in the extremal black hole spacetime.

\begin{figure}
\centering
\includegraphics[width=3in]{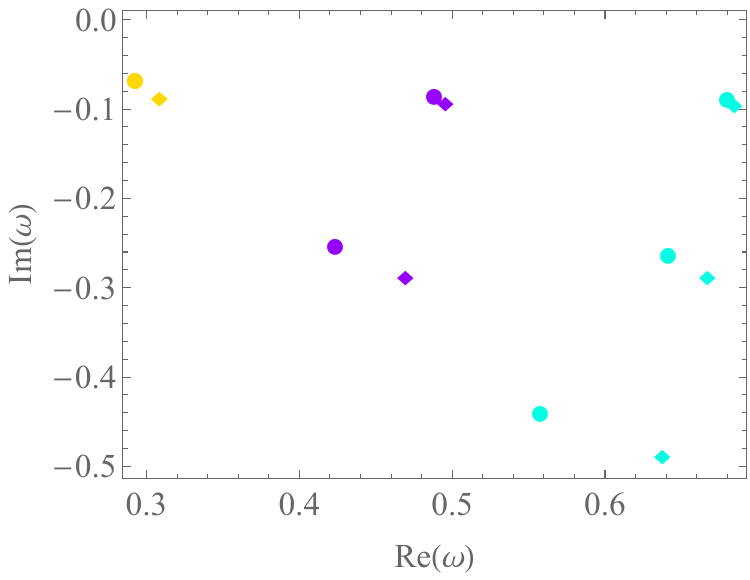}
\caption{(Color online) The QNFs for different $\ell$ and $n$ at $\theta=0.0,~0.2758$. Circles denote $\theta = 0.2758$, and diamonds denote $\theta = 0.0$. Yellow, purple and cyan represent $\ell=1,2,3$, respectively. Points from top to bottom correspond to $n=0,1,2$ ($\mu=0.2$).}
\label{fig:Re_vs_Im}
\end{figure}

Fig.~\ref{fig:Re_vs_Im} displays the behavior of QNFs as functions of $\ell$ and $n$ for fixed $\theta=0.0,~0.2758$ with mass $\mu=0.2$. For a fixed $\theta$, increasing $\ell$ raises and broadens the potential barrier, leading to an increase in the real part and a decrease in the absolute value of the imaginary part $-\mathrm{Im}(\omega)$. In contrast, increasing the overtone number $n$ slightly decreases the real part and significantly increases $-\mathrm{Im}(\omega)$, corresponding to faster decay of higher-overtone modes. These trends are qualitatively consistent with the known behavior of the classical Schwarzschild black hole, indicating that noncommutative corrections do not alter the fundamental dependence of the QNFs with massive scalar field.

\begin{figure*}
\centering
\subfigure[~$\ell=1,n=0$]
{\includegraphics[width=1.7in]{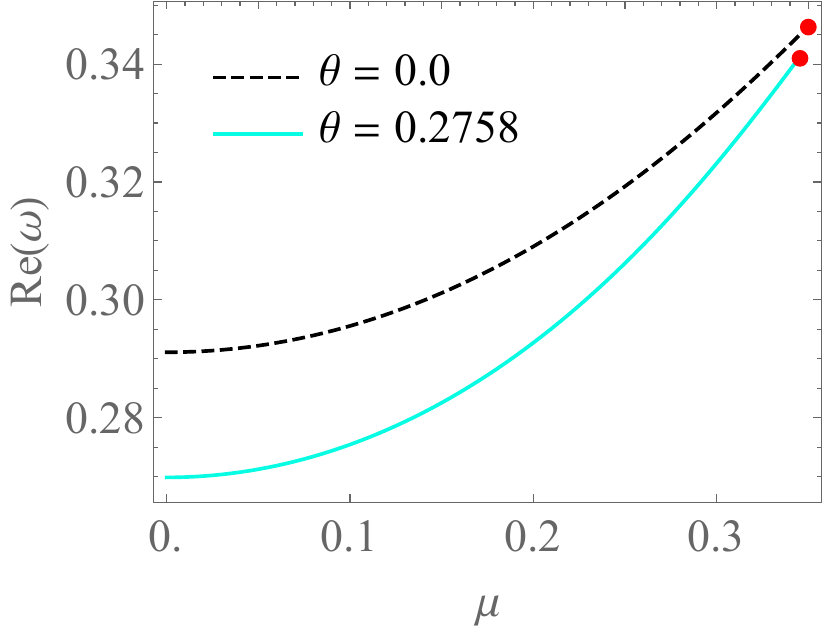}}
\subfigure[~$\ell=1,n=0$]
{\includegraphics[width=1.7in]{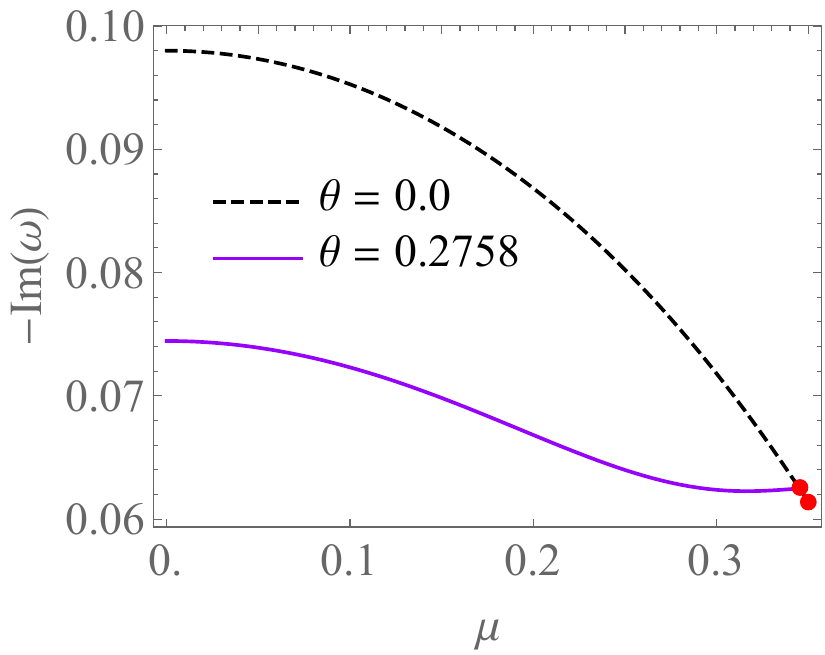}}
\subfigure[~$\ell=2,n=0$]
{\includegraphics[width=1.7in]{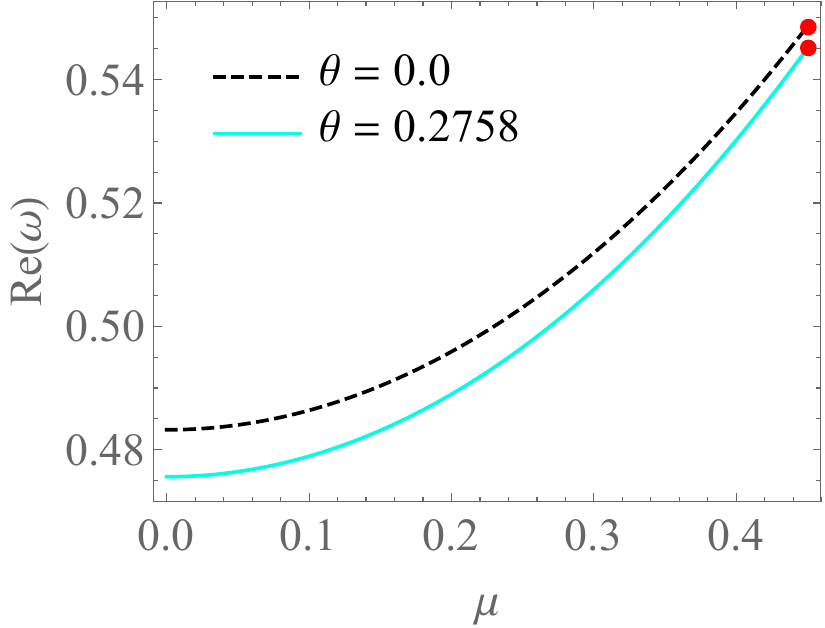}}
\subfigure[~$\ell=2,n=0$]
{\includegraphics[width=1.7in]{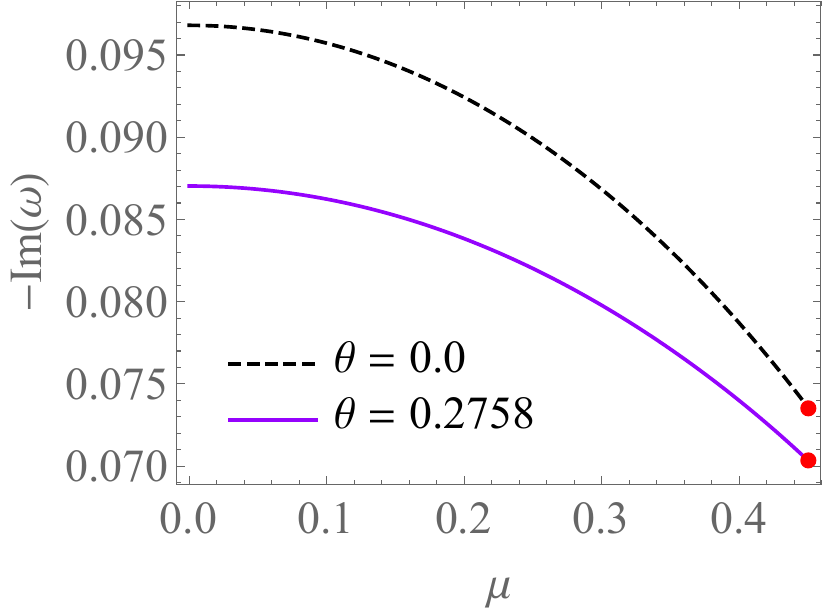}}\\
\subfigure[~$\ell=2,n=1$]
{\includegraphics[width=1.7in]{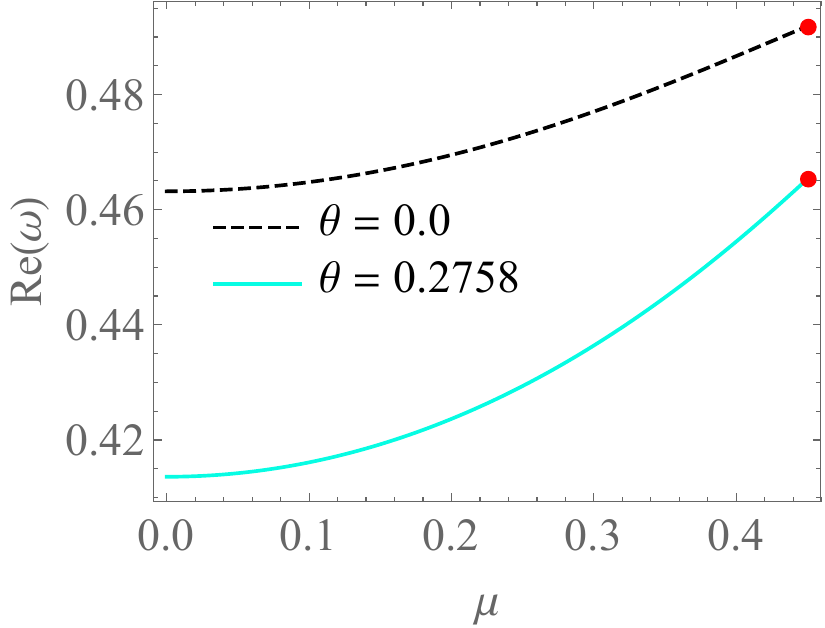}}
\subfigure[~$\ell=2,n=1$]
{\includegraphics[width=1.7in]{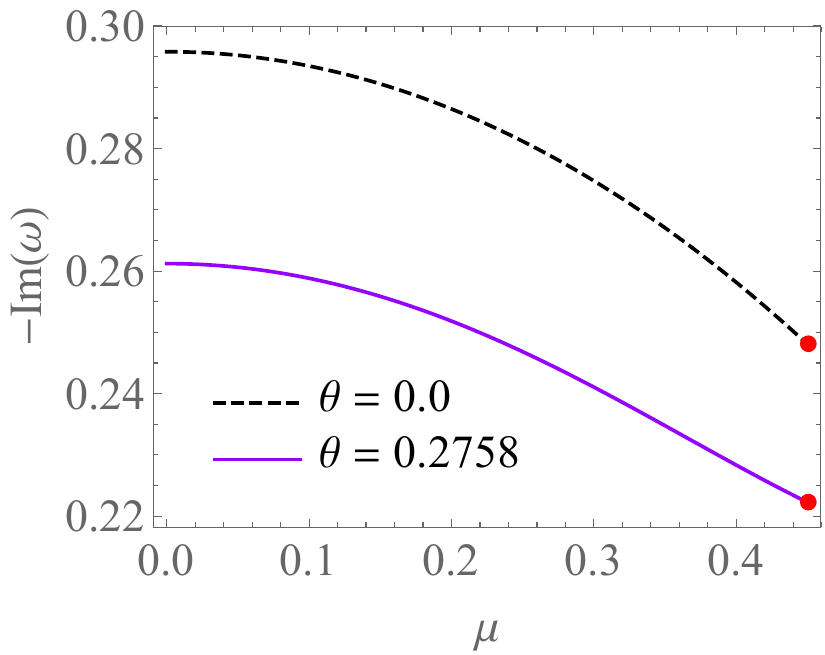}}
\subfigure[~$\ell=3,n=0$]
{\includegraphics[width=1.7in]{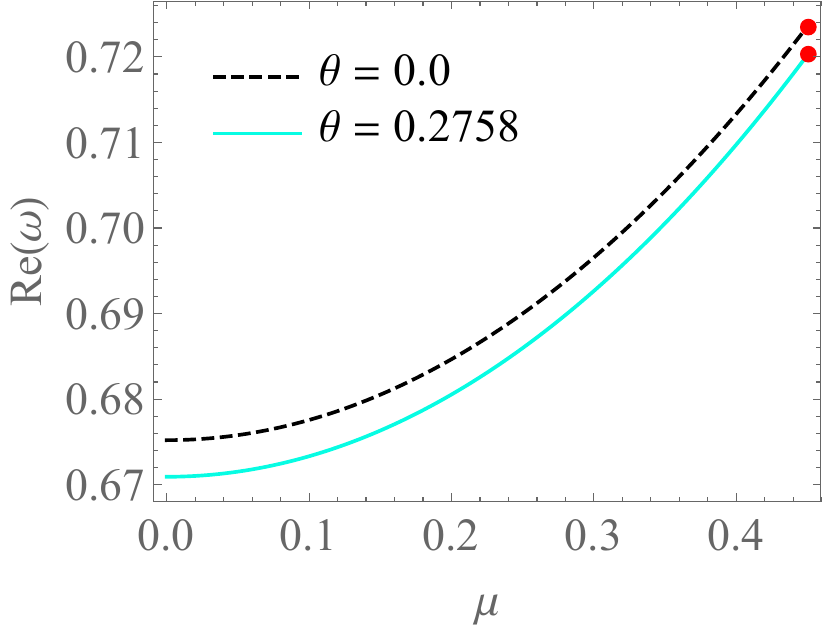}}
\subfigure[~$\ell=3,n=0$]
{\includegraphics[width=1.7in]{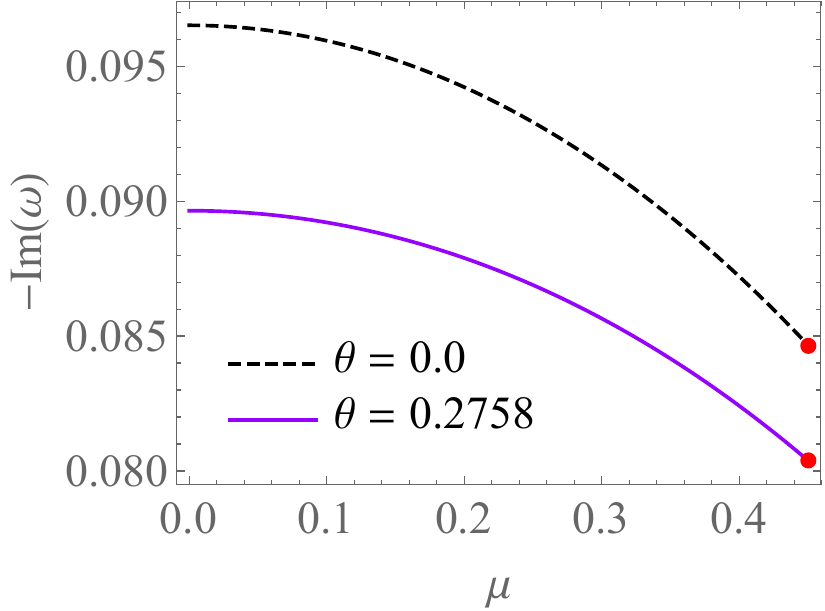}}\\
\subfigure[~$\ell=3,n=1$]
{\includegraphics[width=1.7in]{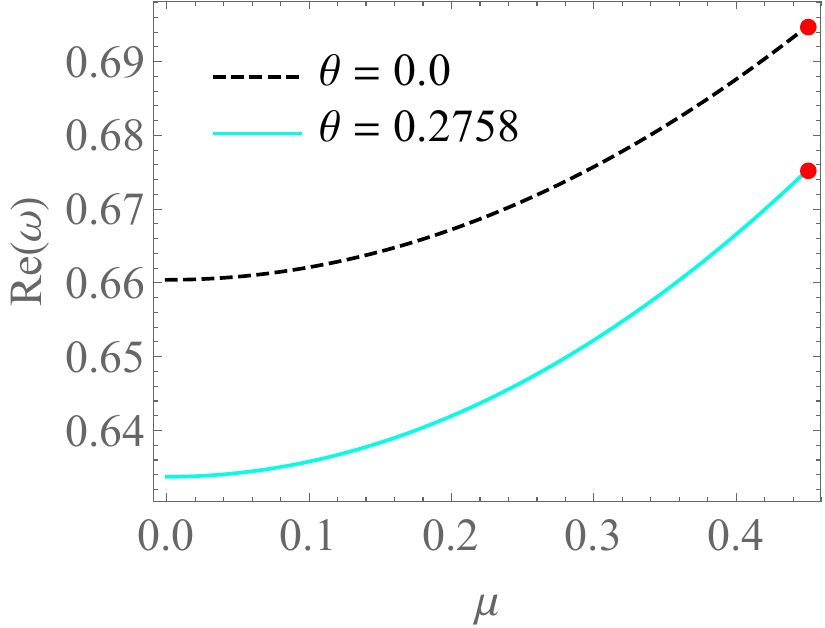}}
\subfigure[~$\ell=3,n=1$]
{\includegraphics[width=1.7in]{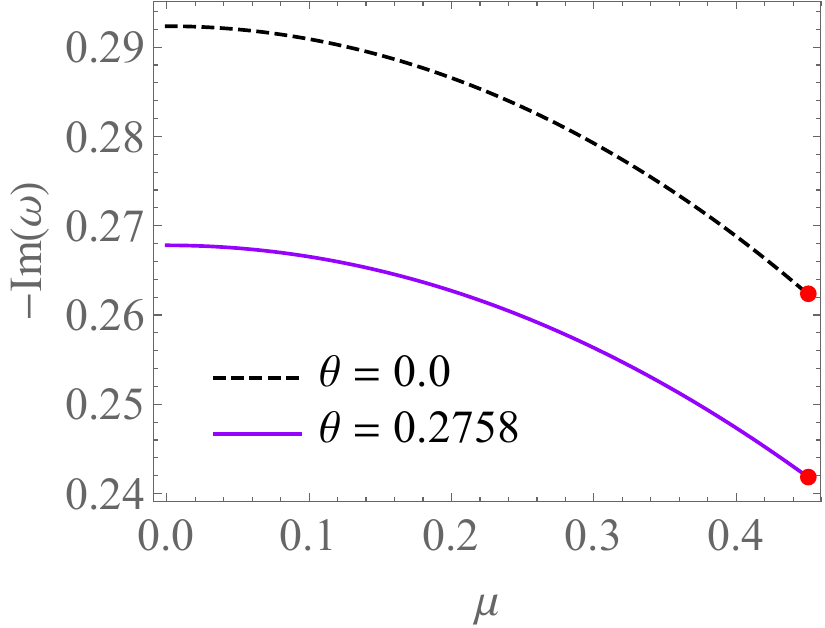}}
\subfigure[~$\ell=3,n=2$]
{\includegraphics[width=1.7in]{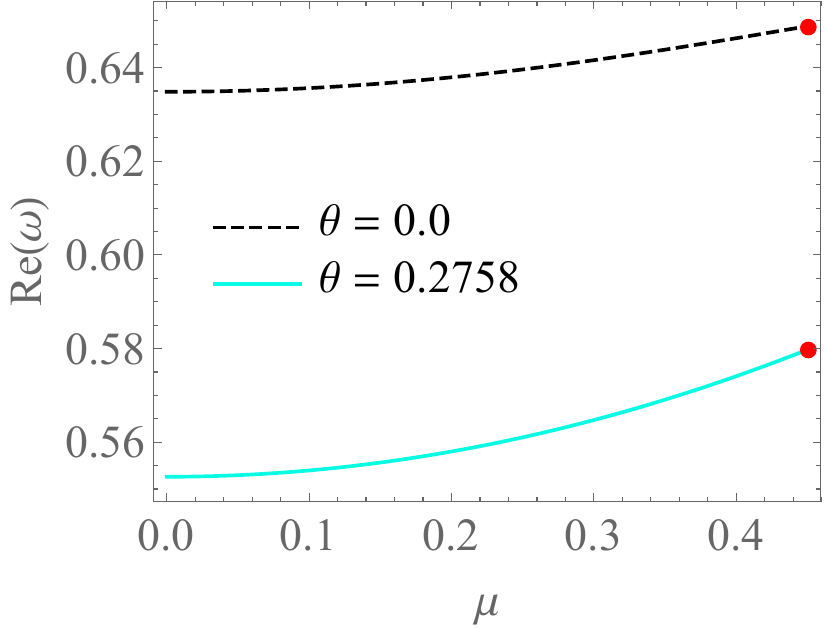}}
\subfigure[~$\ell=3,n=2$]
{\includegraphics[width=1.7in]{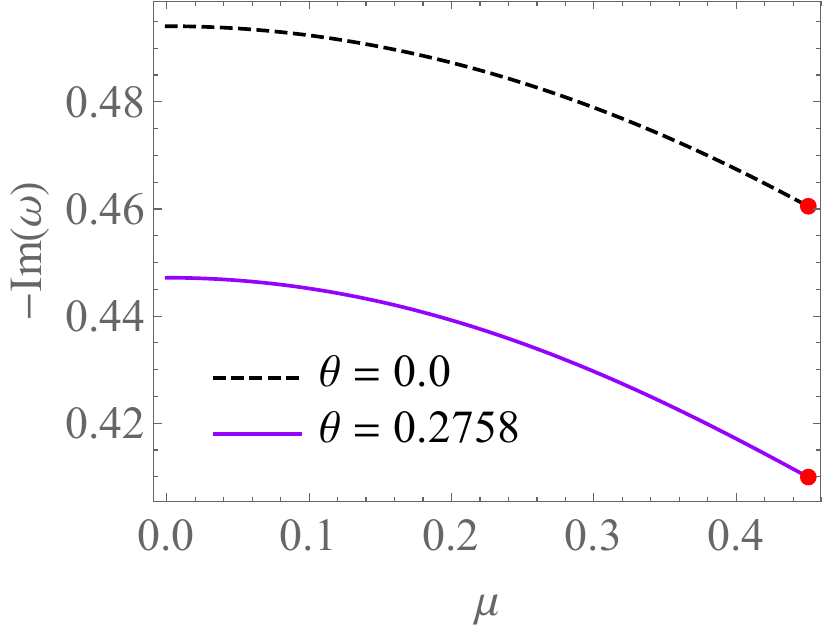}}
\caption{(Color online) The evolution of the real and imaginary parts of the QNFs—$\mathrm{Re}(\omega)$ and $\mathrm{Im}(\omega)$—as a function of $\mu$ for classical Schwarzschild black hole $\theta=0.0$ and extreme black hole $\theta=0.2758$.}
\label{fig:ReIm_vs_m_without_time}
\end{figure*}

To gain a clearer understanding of the regulatory effects of mass on QNFs, we focus on the extremal black hole case $\theta=0.2758$, and comparing the results with those of classical Schwarzschild black hole ($\theta=0.0$). Fig.~\ref{fig:ReIm_vs_m_without_time} illustrates the QNFs maintain $\mathrm{Im}(\omega)<0$ throughout the entire mass range of the scalar field under consideration, demonstrating that the scalar field mass does not destabilize the black hole. The modulation effect of $\mu$ on the QNFs can be summarized as follows. First, the real part $\mathrm{Re}(\omega)$ increases monotonically with $\mu$, which reflects the increase in the effective potential height with $\mu$, as shown in Fig.~\ref{fig:V_vs_rstar_without_time}(c), leading to higher oscillation frequencies. While the imaginary part $-\mathrm{Im}(\omega)$ decreases monotonically with increasing $\mu$, which means that more massive scalar field perturbations decay more slowly, which can be physically attributed to the higher potential barrier that tends to trap the wave in the potential well, reducing energy radiation to infinity. Moreover, in the extremal black hole, the modulation effect of the mass on QNMs is qualitatively similar to that in the classical Schwarzschild black hole, but with quantitative differences: for the same $\mu$, the value of $-\mathrm{Im}(\omega)$ in the extremal black hole is smaller than that in the classical Schwarzschild black hole. This further confirms that noncommutative quantum corrections tend to slow down the decay of perturbations.

Of particular interest is the synergistic effect between the angular quantum number $\ell$ and the scalar field mass $\mu$. As shown in Fig.~\ref{fig:ReIm_vs_m_without_time}(a)-(b), for $\ell=1$ and in the region of relatively large physically allowed masses, the QNFs of the extremal black hole are very close to those of the classical Schwarzschild black hole. This phenomenon suggests that for low angular momentum modes, the mass effect of the scalar field partially masks the quantum corrections from noncommutative geometry, making the black hole response more reminiscent of the classical expectation. This finding may has potential implications for understanding the radiation behavior of massive scalar field during primordial black hole evaporation.

To summarize, in this section we examine for the first time the regulatory effect of scalar field mass on the QNFs of a NCG-Schwarzschild BH, which paves the way for subsequent investigations of the GFs and ACS.

\section{Greybody factors}\label{secV}
In black hole physics, the greybody factors (GFs) describe the transmission probability of a quantum field through the potential barrier surrounding a black hole. The GFs essentially define the transmission coefficient from the horizon to infinity, $\Gamma(\Omega)\equiv|T(\Omega)|^2$, as a function of the real frequency $\Omega$. The GFs scattering have the following boundary conditions~\cite{Chen2025CPC,Zhang2026CTP,Becar2026arXiv,Ma2024EPJP,Arbelaez2026arXiv,Konoplya2019CQG}:
\begin{eqnarray}
\psi=
\begin{cases}
T e^{-i\Omega r_*},\quad r_*\to -\infty,\\
e^{-i\Omega r_*}+R e^{i\Omega r_*},\quad r_*\to +\infty,
\end{cases}
\end{eqnarray}
where $T$ and $R$ represent the transmission coefficient and reflection coefficient, respectively. According to normalization, $T$ and $R$ are related by
\begin{eqnarray}
|T|^2+|R|^2=1.
\end{eqnarray}
In this study, we employ the third-order WKB approximation approach to compute the transmission coefficients. This method expands the wave equation around the peak of the potential barrier and matches asymptotic solutions near the horizon and at infinity, yielding the scattering coefficients in the form
\begin{eqnarray}
|R|^2&=&\frac{1}{1+e^{-2\pi i\mathcal{K}(\Omega)}},\\
\Gamma(\Omega)&=&|T|^2=1-|R|^2,
\end{eqnarray}
where $\mathcal{K}$ is a parameter which can be obtained by the WKB formula~\cite{Konoplya2011RMP},
\begin{eqnarray}
\mathcal{K}=\frac{i Q_0}{\sqrt{2Q_0''}}+\sum_{i=2}^3\Lambda_i,
\end{eqnarray}
where \( Q_0 = \Omega^2 - V(r_*^{(0)}) \), \( V(r_*^{(0)}) \) is the maximum of the effective potential, and \( \Lambda_i \) are correction terms associated with the WKB order.

\begin{figure*}
\centering
\subfigure[~$\ell=1,\mu=0.0$]
{\includegraphics[width=2.2in]{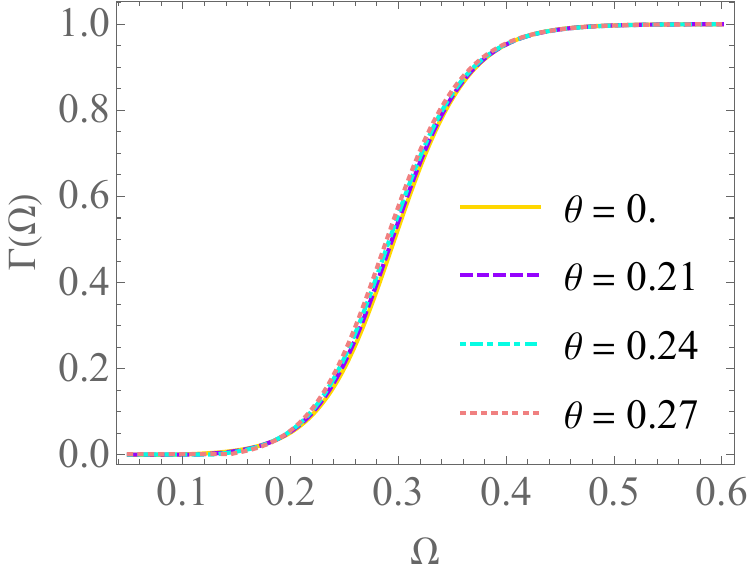}}
\subfigure[~$\ell=2,\mu=0.0$]
{\includegraphics[width=2.2in]{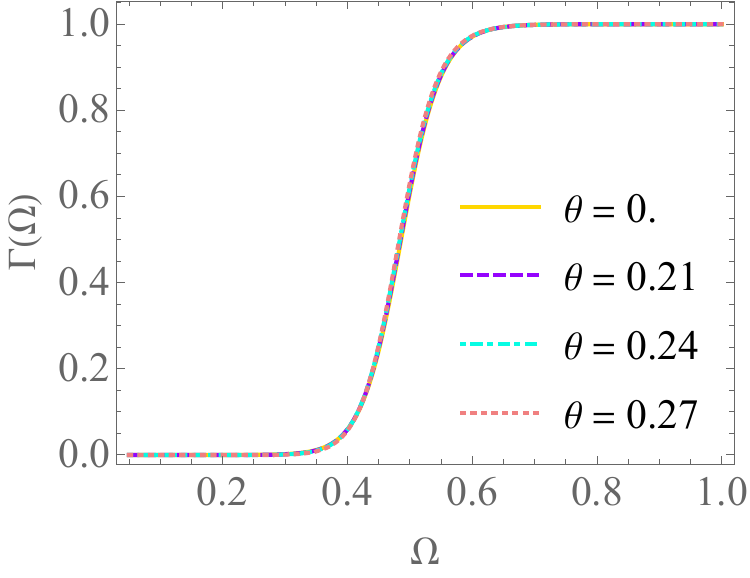}}
\subfigure[~$\ell=3,\mu=0.0$]
{\includegraphics[width=2.2in]{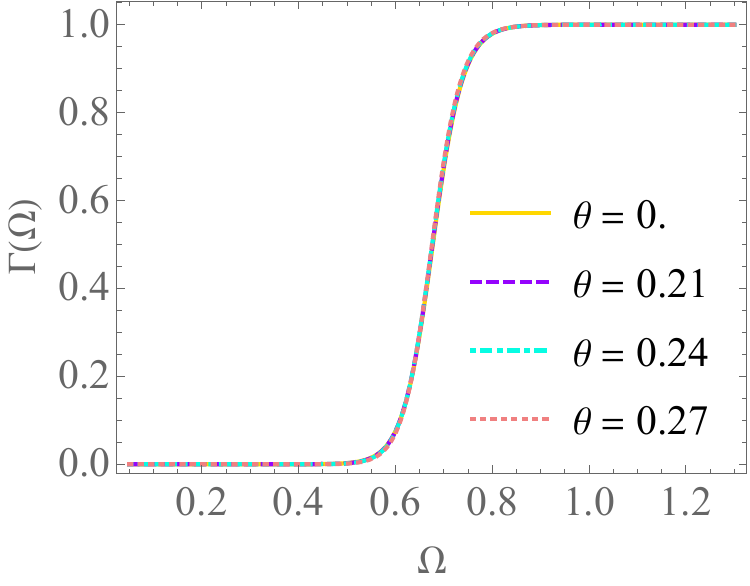}}\\
\subfigure[~$\ell=1,\theta=0.16$]
{\includegraphics[width=2.2in]{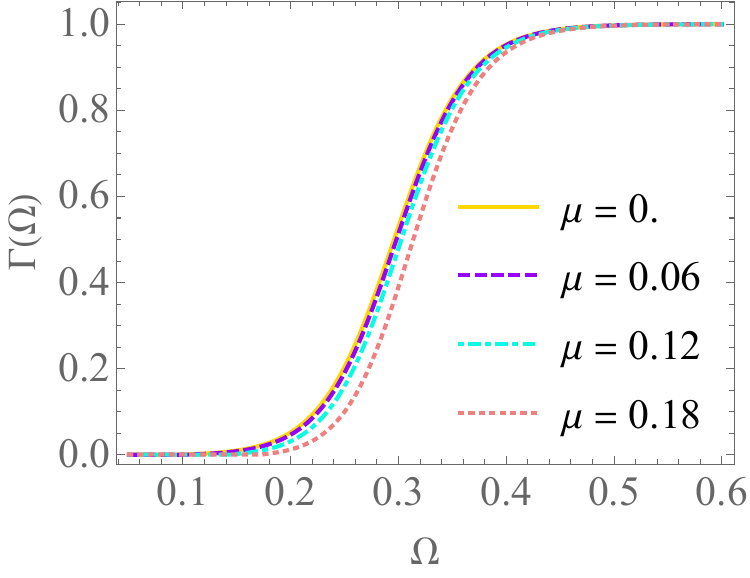}}
\subfigure[~$\ell=2,\theta=0.16$]
{\includegraphics[width=2.2in]{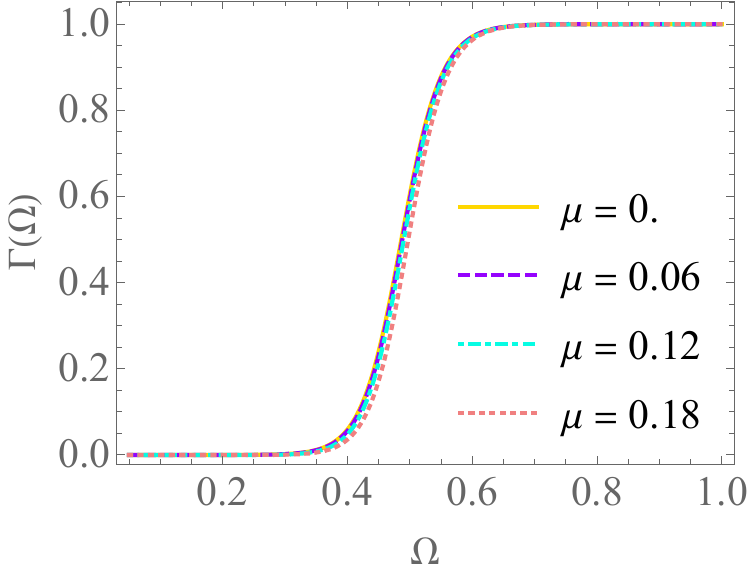}}
\subfigure[~$\ell=3,\theta=0.16$]
{\includegraphics[width=2.2in]{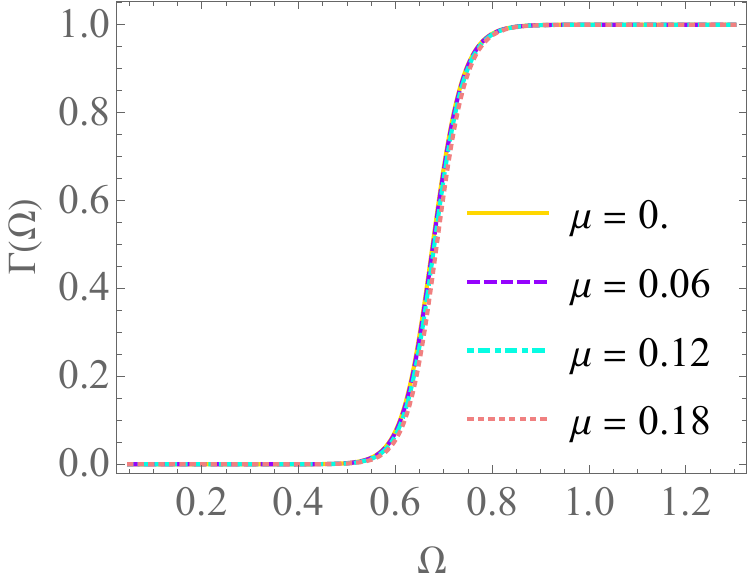}}
\caption{(Color online) The evolution of the GFs as a function of frequency with representative $\theta$ and $\mu$.}
\label{fig:Gamma_vs_Omega}
\end{figure*}

Fig.~\ref{fig:Gamma_vs_Omega} displays the GFs as a function of the frequency $\Omega$, with the first row fixing the scalar field mass $\mu=0.0$ and the second row fixing the noncommutative parameter $\theta=0.16$, for angular quantum numbers $\ell=1,2,3$. Overall, the GFs exhibit that they approach zero in the low-frequency regime, and tend to unity in the high-frequency regime. This indicates that low-frequency waves are almost completely reflected by the potential barrier, while high-frequency waves penetrate the barrier with negligible reflection. This behavior is consistent with the “filtering” property of the effective potential—modes with frequencies below the barrier height have difficulty escaping.

For fixed $\theta,~\mu$, increasing the angular quantum number $\ell$ shifts the GFs curve to the right, meaning that higher frequencies are required to achieve the same transmission probability. This is because a larger $\ell$ corresponds to a higher and broader effective potential barrier (see Fig.~\ref{fig:V_vs_rstar_without_time}(a)), making it more difficult for particles to tunnel through. As shown in Table~\ref{tab:combined_Gamma}, for $\mu=0.0$ and $\theta=0.21$ as an instant, the frequency at which $\Gamma=0.5$ is approximately $\Omega=0.294067$ for $\ell=1$, increases to $\Omega=0.485067$ for $\ell=2$, and further to $\Omega=0.676467$ for $\ell=3$.

\begin{table}
\centering
\caption{Frequencies $\Omega_m$ for the GFs $\Gamma=0.5$ under different parameters $\theta$ and scalar field masses $\mu$.}
\begin{tabular}{|c|ccc|}
\hline\hline
\multirow{1}{*}{$\Omega_m$} & \multicolumn{1}{c}{$\ell=1$} & \multicolumn{1}{c}{$\ell=2$} & \multicolumn{1}{c|}{$\ell=3$} \\
\hline
\multicolumn{4}{|c|}{$\mu=0.0$} \\
\hline
$\theta=0.00$   & 0.296267 & 0.486067 & 0.6772 \\
$\theta=0.21$  & 0.294067 & 0.485067 & 0.676467 \\
$\theta=0.24$  & 0.291667 & 0.484067 & 0.675933 \\
$\theta=0.27$  & 0.289067 & 0.483 & 0.675467 \\
\hline
\multicolumn{4}{|c|}{$\theta=0.16$} \\
\hline
$\mu=0.00$ & 0.295868 & 0.486404 & 0.676931\\
$\mu=0.06$ & 0.298113 & 0.487957 & 0.677926\\
$\mu=0.12$ & 0.302674 & 0.490379 & 0.680453\\
$\mu=0.18$ & 0.313675 & 0.497424 & 0.685477\\
\hline\hline
\end{tabular}
\label{tab:combined_Gamma}
\end{table}

The noncommutative parameter $\theta$ also quantitatively affects the GFs. Specifically, as seen in the first row of Fig.~\ref{fig:Gamma_vs_Omega} and Table~\ref{tab:combined_Gamma}, for fixed $\mu=0.0$, increasing $\theta$ shifts the GFs curve to the left (toward lower frequencies). This implies that, at a given frequency, a black hole with a larger noncommutative parameter has a larger GF, i.e., particles are more likely to tunnel through the potential barrier. This phenomenon can be attributed to the reduction of the effective potential barrier height with increasing $\theta$ (see Fig.~\ref{fig:V_vs_rstar_without_time}(b)), which in turn reduces the reflection probability. Notably, this effect is more pronounced for low angular quantum numbers; for large $\ell$, the curves for different $\theta$ gradually converge, indicating that quantum corrections have a weaker influence on high-$\ell$ modes.

The second row of Fig.~\ref{fig:Gamma_vs_Omega}, with fixed $\theta=0.16$, illustrates the modulation of the GFs by the scalar field mass $\mu$. The overall trend is that increasing $\mu$ shifts the GFs curve to the right, requiring higher frequencies to achieve the same transmission probability. This behavior arises because the effective potential for a massive scalar field asymptotically approaches $\mu^2$ at infinity, and the barrier height increases with $\mu$ (see Fig.~\ref{fig:V_vs_rstar_without_time}(c)), thereby enhancing reflection. Again, this mass effect is more pronounced for smaller $\ell$.

In a charged NCG-Schwarzschild BH, the charge causes the GFs to increase; however, it suppresses the regulatory effect of $\theta$ on the GFs, so that $\theta$ has virtually no influence on the GFs~\cite{Ma2024EPJP}. This may reflect different regulatory mechanisms of the effective potential by charge and noncommutative parameters.

\section{Absorption cross section}\label{secVI}
With the GFs in hand, we can easily calculate the absorption cross section (ACS). The ACS characterizes the effective area over which a black hole absorbs energy from incident waves and is a fundamental quantity in black hole scattering theory, and is given by~\cite{Fan2025arXiv,Becar2026arXiv,Tang2026arXiv}
\begin{eqnarray}
\sigma=\sum_{\ell=0}^\infty \sigma_\ell,\label{eq:total_cs}
\end{eqnarray}
where each partial contribution is computed from the transmission coefficient through
\begin{eqnarray}
\sigma_\ell=\frac{\pi}{\Omega^2}(2\ell+1)\Gamma_\ell(\Omega),\label{eq:cs_ell}
\end{eqnarray}
with $\Omega$ the frequency of the incident wave, $\Gamma_\ell(\Omega)$ is the GFs for the $\ell$-th partial wave. The factor $(2\ell+1)$ arises from the degeneracy of angular momentum states, while $\pi/\Omega^2$ originates from the normalization of the partial wave expansion of a plane wave. 

\begin{figure*}
\centering
\subfigure[~$\ell=1,\mu=0.0$]
{\includegraphics[width=2.2in]{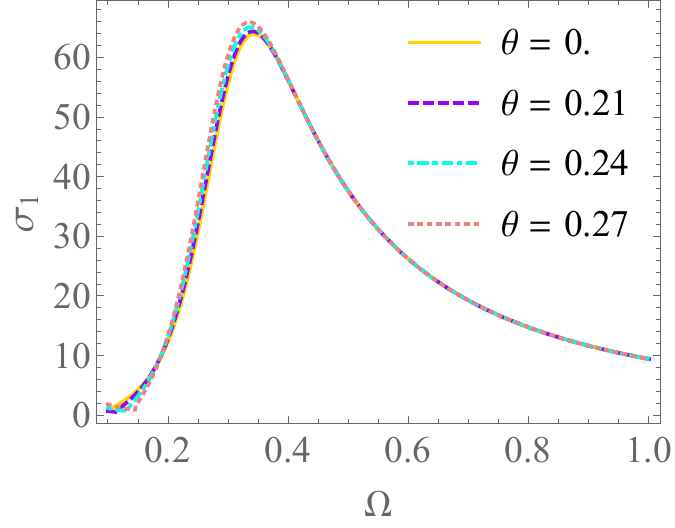}}
\subfigure[~$\ell=2,\mu=0.0$]
{\includegraphics[width=2.2in]{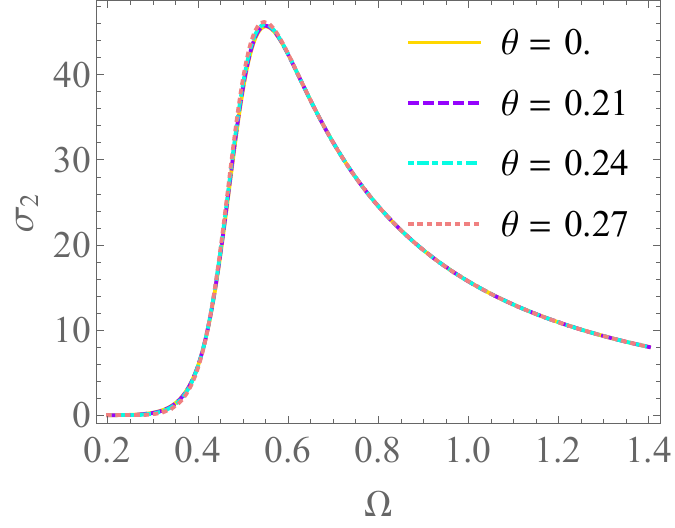}}
\subfigure[~$\ell=3,\mu=0.0$]
{\includegraphics[width=2.2in]{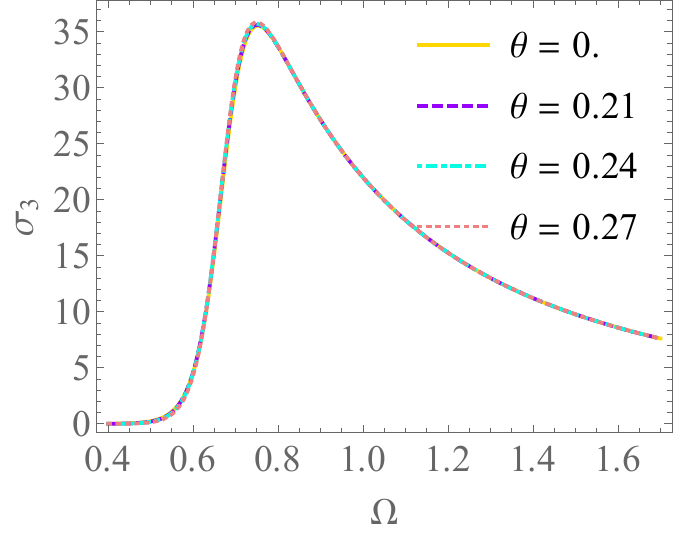}}\\
\subfigure[~$\ell=1,\theta=0.16$]
{\includegraphics[width=2.2in]{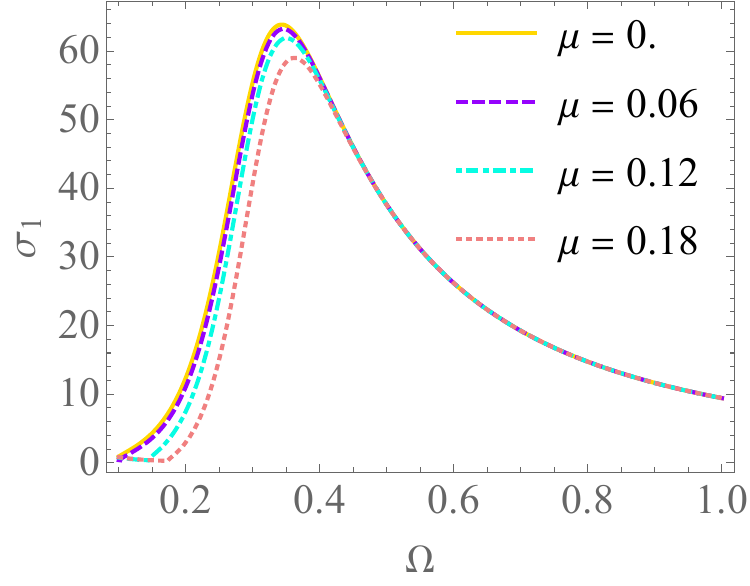}}
\subfigure[~$\ell=2,\theta=0.16$]
{\includegraphics[width=2.2in]{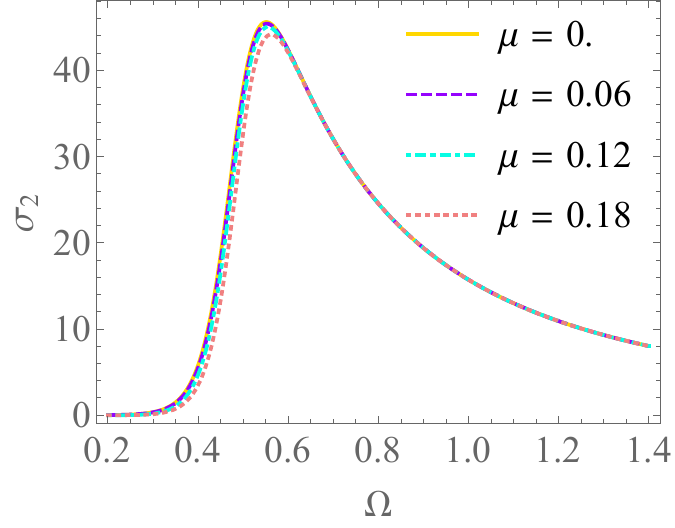}}
\subfigure[~$\ell=3,\theta=0.16$]
{\includegraphics[width=2.2in]{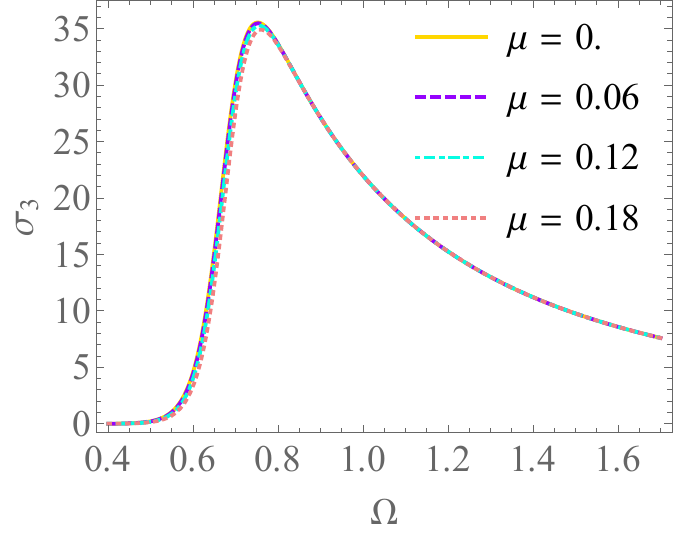}}
\caption{(Color online) The evolution of the ACS as a function of frequency with representative $\theta$ and $\mu$.}
\label{fig:cs_vs_Omega}
\end{figure*}

Based on the GFs computed in the previous section, we calculate the ACS of NCG-Schwarzschild BH using Eqs.~(\ref{eq:total_cs})-(\ref{eq:cs_ell}). Fig.~\ref{fig:cs_vs_Omega} displays the partial wave ACS $\sigma_\ell$ as a function of frequency $\Omega$. Firstly, the ACS increases with frequency at the beginning, until reaches a maximum, and then gradually decreases accompanied by oscillations. This behavior reflects the scattering characteristics of the effective potential: in the intermediate frequency regime, where the wavelength is comparable to the black hole size, resonant effects are most prominent, leading to a peak in the ACS. As the frequency increases further, the ACS tends to stabilize, and the contributions from different partial waves become distinct.

\begin{table*}
\centering
\caption{Peak values of the wave-division ACS and their corresponding frequencies for different parameters $\theta$ and scalar field mass $\mu$.}
\begin{tabular}{|c|cc|cc|cc|}
\hline\hline
\multirow{2}{*}{Parameters} & \multicolumn{2}{c|}{$\ell=1$} & \multicolumn{2}{c|}{$\ell=2$} & \multicolumn{2}{c|}{$\ell=3$} \\
\cline{2-7}
& $\sigma_{\max}$ & $\Omega_c$ & $\sigma_{\max}$ & $\Omega_c$ & $\sigma_{\max}$ & $\Omega_c$ \\
\hline
\multicolumn{7}{|c|}{$\mu=0.0$} \\
\hline
$\theta=0.00$   & 63.9186 & 0.343    & 45.6888 & 0.550267 & 35.5408 & 0.752533 \\
$\theta=0.21$  & 64.3608 & 0.3408   & 45.7675 & 0.549733 & 35.6194 & 0.751267 \\
$\theta=0.24$  & 65.0828 & 0.3378   & 45.9368 & 0.548533 & 35.7009 & 0.750467 \\
$\theta=0.27$  & 65.9325 & 0.334867 & 46.2097 & 0.546867 & 35.8299 & 0.7494   \\
\hline
\multicolumn{7}{|c|}{$\theta=0.16$} \\
\hline
$\mu=0.00$   & 63.8688 & 0.343133 & 45.6122 & 0.5512   & 35.5594 & 0.752133 \\
$\mu=0.06$  & 63.2233 & 0.3456   & 45.4198 & 0.552733 & 35.4853 & 0.753067 \\
$\mu=0.12$  & 61.8842 & 0.350267 & 45.0368 & 0.555    & 35.2808 & 0.755467 \\
$\mu=0.18$  & 59.0071 & 0.361933 & 44.166  & 0.5618   & 34.9161 & 0.7602   \\
\hline\hline
\end{tabular}
\label{tab:combined_absorption}
\end{table*}

The parameters $\ell,~\theta$, and $\mu$ quantitatively influence the evolution of the ACS. First, for fixed $\mu$ and $\theta$, the partial wave ACS for different $\ell$ exhibits distinct threshold behavior. As shown in Table~\ref{tab:combined_absorption}, a larger angular quantum number $\ell$ leads to a higher peak frequency $\Omega_c$ and a smaller peak value $\sigma_\mathrm{max}$. This is because a higher centrifugal barrier shifts the resonance to higher frequencies, while the reduced transmission probability lowers the peak ACS. Second, when $\mu=0.0$ is fixed, increasing $\theta$ increases the peak value of the ACS and shifts the peak position toward lower frequencies. This trend is consistent with the behavior of the GFs: increasing $\theta$ lowers the effective potential barrier, making it easier for particles to penetrate and thereby enhancing absorption. For example, for $\ell=1$, the peak ACS for $\theta=0.27$ is approximately 65.93, significantly higher than the classical Schwarzschild case ($\theta=0.0$) with a value of 63.92. On the other hand, when $\theta=0.16$ is fixed, Table~\ref{tab:combined_absorption} shows that increasing $\mu$ reduces the peak value of the ACS and shifts the peak position toward higher frequencies. This is due to the higher effective potential barrier for massive scalar field, which suppresses the absorption of low-energy particle. For instance, for $\ell=1$, the peak cross section for $\mu=0.18$ drops to 59.01, substantially lower than the massless case value of 63.87. This result is fully consistent with the mass modulation effect observed in the GFs.

It is noteworthy that the noncommutative parameter $\theta$ and the scalar field mass $\mu$ have opposite effects on the ACS: increasing $\theta$ enhances absorption, while increasing $\mu$ suppresses it. In the context of primordial black hole evaporation, these two effects may compete with each other, jointly determining the final radiation spectrum. For light scalar fields (small $\mu$), noncommutative quantum corrections may significantly enhance low-frequency radiation in the extremal black hole regime. For heavy scalar fields (large $\mu$), the mass effect dominates, and the ACS approaches the classical result. This phenomenon shares a similar physical origin with QNMs results at $\ell=1$.

\section{Conclusion and outlook}\label{secVII}
In this study, we systematically investigate massive scalar field perturbations in a noncommutative-geometry-inspired Schwarzschild black hole (NCG-Schwarzschild BH), with a focus on the modulation effects of the noncommutative parameter \(\theta\) and the scalar field mass \(\mu\) on the quasinormal modes frequencies (QNFs), greybody factors (GFs), and absorption cross section (ACS). Based on the NCG-Schwarzschild BH featuring a Gaussian matter distribution, we derive the Klein-Gordon equation for a massive scalar field in curved spacetime and, by introducing the tortoise coordinate, cast it into the standard form of a \((1+1)\)-dimensional wave equation, presenting an explicit expression for the effective potential. Using the third-order WKB approximation approach, we systematically examine the QNFs for \(\ell > n\) under different values of \(\theta\) and \(\mu\), as well as the behavior of the GFs and ACS.

The QNFs of a NCG-Schwarzschild BH in the case \(\mu \neq 0\) is studied for the first time. In qualitative agreement with the massless perturbation results, all QNFs satisfy \(\operatorname{Im}(\omega) < 0\), confirming that the NCG-Schwarzschild BH is stable under massive scalar field perturbations. Moreover, an increase in the scalar field mass \(\mu\) leads to an increase in the real part of the frequency and a decrease in the absolute value of the imaginary part, i.e., the perturbations decay more slowly. Notably, when \(\ell = 1\) and \(\mu\) is sufficiently large, the QNFs of the extremal black hole approach those of the classical Schwarzschild black hole, indicating a partial cancellation between the scalar field mass effect and the noncommutative quantum corrections in the low-angular-momentum regime.

The GFs increase with frequency in a S-shaped curve, reflecting the filtering property of the effective potential. Increasing \(\theta\) shifts the GFs curve toward lower frequencies, facilitating particle penetration through the potential barrier, whereas increasing \(\mu\) shifts the curve toward higher frequencies, enhancing reflection. These trends are more pronounced for low angular quantum numbers, indicating that noncommutative corrections have a relatively weaker effect on large-\(\ell\) modes. The partial-wave ACS, computed from the GFs, first increase with frequency, reach a peak, and then gradually decrease. Larger \(\ell\) leads to a higher peak frequency and a smaller peak amplitude. Increasing \(\theta\) increases the peak ACS and shifts it toward lower frequencies, while increasing \(\mu\) decreases the peak and shifts it toward higher frequencies. The opposite effects of these two types of parameters on the ACS may give rise to competitive behavior during the evaporation of primordial black holes, jointly determining the final form of the radiation spectrum.

The findings of this paper have theoretical significance for understanding black hole perturbation dynamics in the presence of quantum gravity effects and also provide a necessary theoretical foundation for predicting the energy spectrum of primordial black hole evaporation products. In future work, we intend to extend this study along two main directions, including cross-validating the results in the extreme regime using spectral methods or the Pad\'e-improved WKB approximation approach, and exploring massive perturbation fields with spin \(s=1,2\) to reveal the competition and synergy between field mass and quantum corrections. These efforts are expected to provide a useful reference for subsequent theoretical and experimental investigations in this field.

\appendix
\renewcommand{\thesection}{Appendix \Alph{section}}
\section{Results of the pure 6th-order WKB approximation approach}\label{compare}
The WKB approximation approach is one of the commonly employed semianalytical methods for computing QNMs~\cite{Iyer1987PRDI,Matyjasek2017PRD,Konoplya2003PRD}. This method is based on the Schr\"odinger-type wave equation, expands the effective potential around the peak of the potential barrier, and matches the asymptotic solutions near the horizon and at infinity to derive an eigenvalue condition for the complex frequency, from which the QNFs of the black hole are obtained.

Depending on the order of expansion, the WKB approximation approach can achieve varying degrees of accuracy. In general, higher-order WKB approximation approach can provide greater precision when the shape of the potential barrier is regular. However, their convergence is sensitive to the behavior of higher-order derivatives of the effective potential at the extremum~\cite{Matyjasek2017PRD,Konoplya2003PRD}. For the NCG-Schwarzschild BH, the effective potential varies with the noncommutative parameter $\theta$, as shown in Fig.~\ref{fig:V_vs_rstar_without_time}(b). In particular, when $\theta$ is large (near the extreme black hole), the behavior near the peak of the potential barrier may deviate from the standard case. Therefore, assessing the applicability of WKB approximation approach of different orders in this model is of significant methodological importance.

\begin{figure*}[htpb]
\centering
\subfigure[~$\ell=1,~n=0$]
{\includegraphics[width=2.2in]{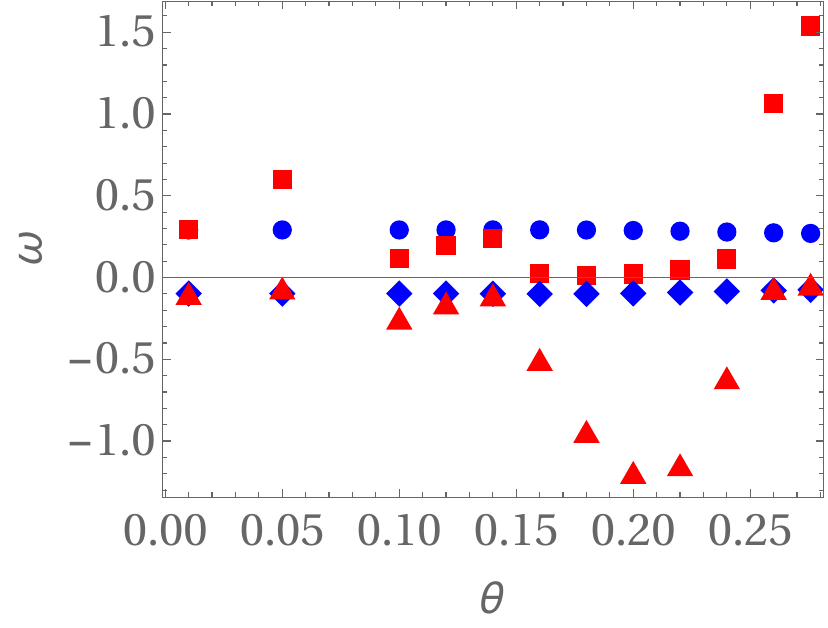}}
\subfigure[~$\ell=2,~n=0$]
{\includegraphics[width=2.2in]{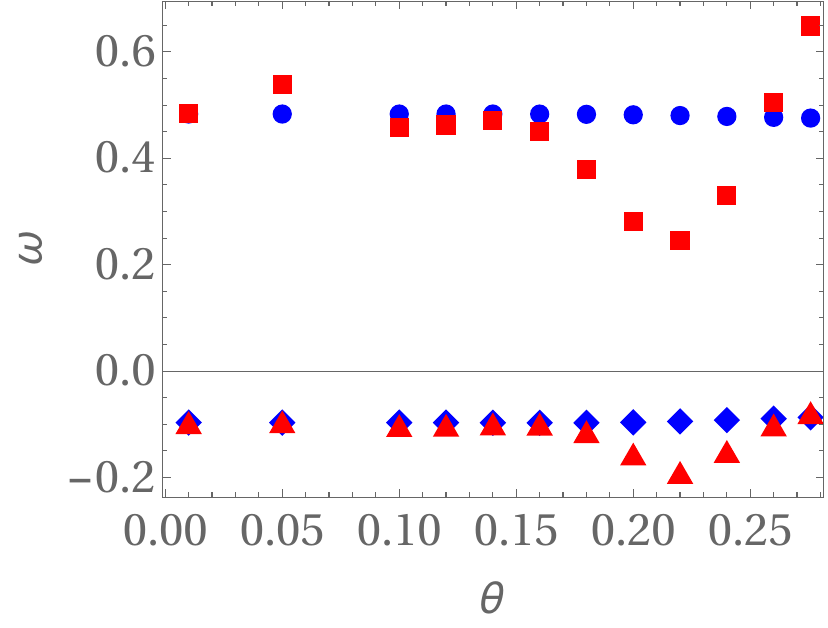}}
\subfigure[~$\ell=2,~n=1$]
{\includegraphics[width=2.2in]{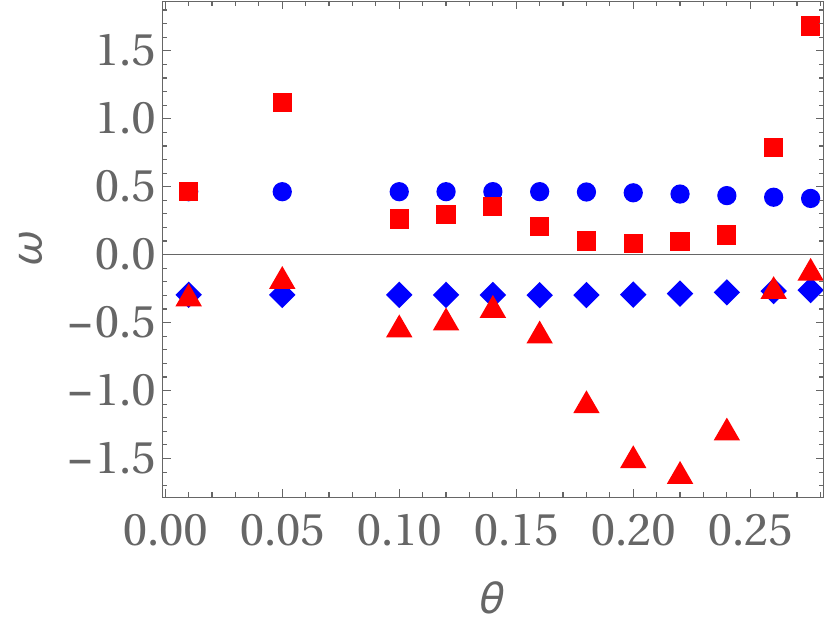}}\\
\subfigure[~$\ell=3,~n=0$]
{\includegraphics[width=2.2in]{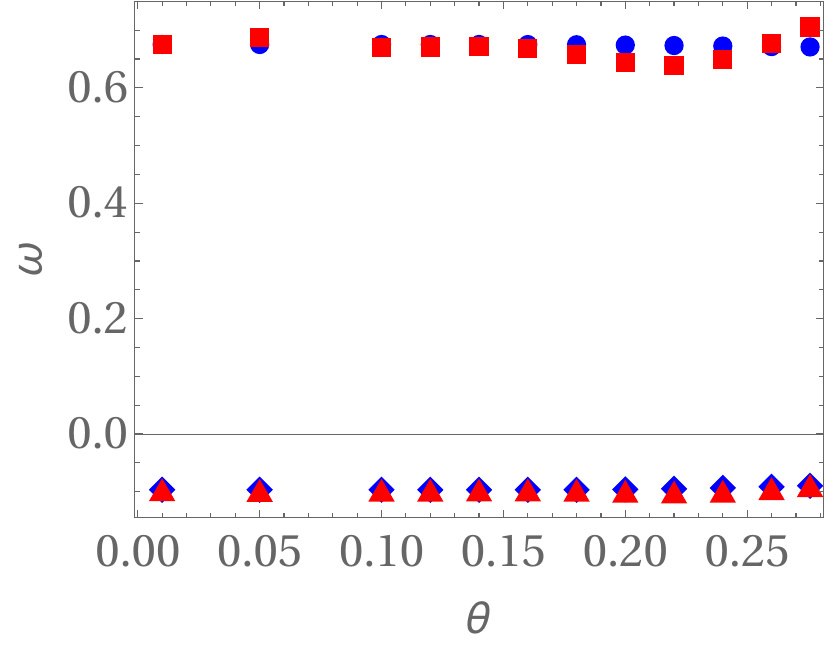}}
\subfigure[~$\ell=3,~n=1$]
{\includegraphics[width=2.2in]{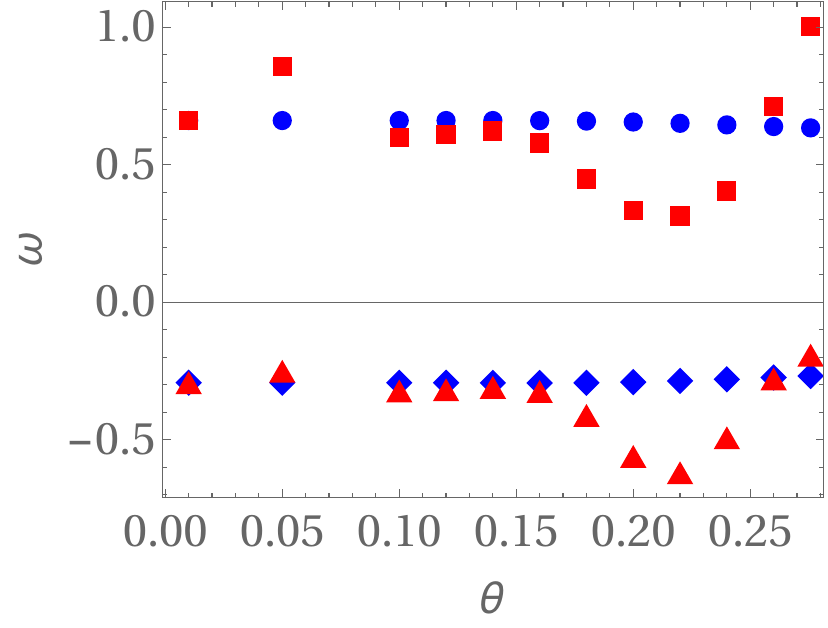}}
\subfigure[~$\ell=3,~n=2$]
{\includegraphics[width=2.2in]{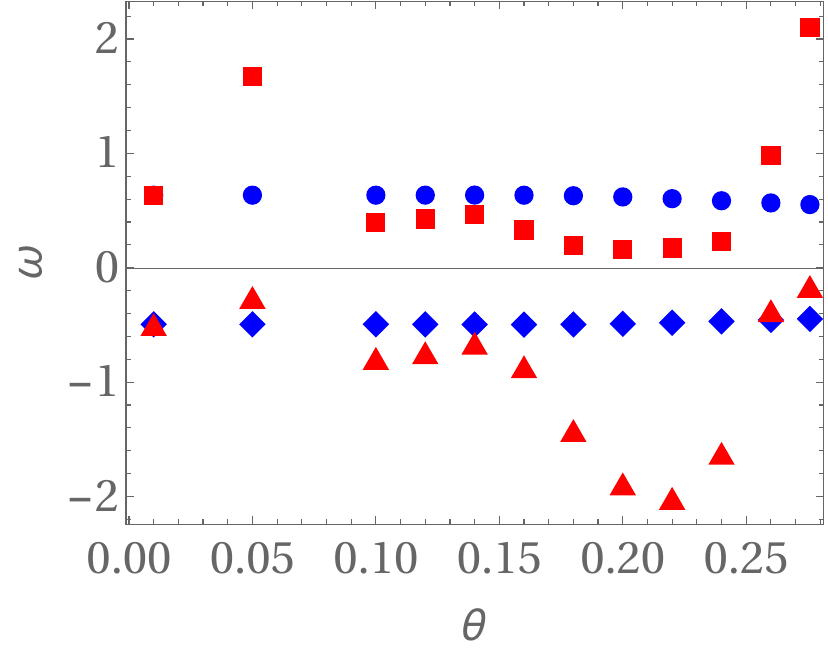}}
\caption{(Color online) Illustration of $\mathrm{Re}(\omega)$ and $\mathrm{Im}(\omega)$ versus $\theta$ with 3rd- and 6th-order WKB approximation approach. The blue circle and diamond represent $\mathrm{Re}(\omega)$ and $\mathrm{Im}(\omega)$ in 3rd order, respectively, while the red square and triangle correspond to the 6th order.}
\label{fig:omega_vs_theta_varying_order}
\end{figure*}

As discussed in the literature, when the noncommutative geometric parameter satisfies Eq.~(\ref{eq:theta_range}), corresponding to a black hole mass $M>3.6M_P$, the third-order WKB approximation yields stable convergent results~\cite{Liang2018CPL,Liang2018CPL2}. However, Batic et al. reported instabilities in the black hole within the mass range $1.91M_P<M<2.3897M_P$, while spectral methods produced reliable results in this regime~\cite{Batic2019EPJP,Batic2024EPJC}. This motivates us to explore whether higher-order WKB approximation can achieve improved accuracy for $M>3.6M_P$. To this end, we preliminarily compute the QNFs using the pure sixth-order WKB approximation and compare them with the third-order results, as shown in Fig.~\ref{fig:omega_vs_theta_varying_order}. The figure illustrates the evolution of the real part $\mathrm{Re}(\omega)$ and $\mathrm{Im}(\omega)$ of the quasinormal mode frequencies with respect to the noncommutative parameter $\theta$ under massless scalar field perturbations.

From Fig.~\ref{fig:omega_vs_theta_varying_order}, we observe that in the region of small $\theta$, i.e., approaching the classical Schwarzschild black hole, the sixth-order WKB approximation approach yields relatively better convergence for both the real and imaginary parts of the frequency, although it remains less than ideal. As $\theta$ increases and approaches its upper bound $\theta=0.2758$ (the extremal black hole), the results from the two methods diverge significantly. The sixth-order WKB approximation approach exhibits unphysical oscillations and even divergent behavior for \(\theta \gtrsim 0.15\), whereas the third-order WKB approximation approach maintains a smooth variation consistent with physical expectations. The numerical instability observed in the pure sixth-order WKB approximation approach suggests that, near the extreme black hole regime, higher-order WKB approximations may suffer from convergence difficulties due to the smallness of the second derivative of the effective potential or the complexity of higher-order derivatives, potentially leading to spurious instability signals. In contrast, the third-order WKB approximation approach demonstrates superior numerical stability in the NCG-Schwarzschild BH, even in the extremal case~\cite{Liang2018CPL,Liang2018CPL2}.

In summary, this constitutes a preliminary test of the WKB approximation approach within the mass range $M>3.6M_P$. For higher-precision verification in extreme cases, future work may draw on results from spectral methods~\cite{Batic2019EPJP,Batic2024EPJC} to cross-validate the Pad\'e-improved WKB approximation approach~\cite{Matyjasek2017PRD,Konoplya2019CQG,Lutfiuoglu2025EPJC}, which is also one of the directions for our future research.

\begin{acknowledgments}
This work was supported by Natural Science Foundation of Jiangsu Province (grant no. BK20220122) and National Natural Science Foundation of China (grant no. 12233002).
\end{acknowledgments}

\end{CJK*}

\begin{thebibliography}{70}%
\makeatletter
\providecommand \@ifxundefined [1]{%
 \@ifx{#1\undefined}
}%
\providecommand \@ifnum [1]{%
 \ifnum #1\expandafter \@firstoftwo
 \else \expandafter \@secondoftwo
 \fi
}%
\providecommand \@ifx [1]{%
 \ifx #1\expandafter \@firstoftwo
 \else \expandafter \@secondoftwo
 \fi
}%
\providecommand \natexlab [1]{#1}%
\providecommand \enquote  [1]{``#1''}%
\providecommand \bibnamefont  [1]{#1}%
\providecommand \bibfnamefont [1]{#1}%
\providecommand \citenamefont [1]{#1}%
\providecommand \href@noop [0]{\@secondoftwo}%
\providecommand \href [0]{\begingroup \@sanitize@url \@href}%
\providecommand \@href[1]{\@@startlink{#1}\@@href}%
\providecommand \@@href[1]{\endgroup#1\@@endlink}%
\providecommand \@sanitize@url [0]{\catcode `\\12\catcode `\$12\catcode
  `\&12\catcode `\#12\catcode `\^12\catcode `\_12\catcode `\%12\relax}%
\providecommand \@@startlink[1]{}%
\providecommand \@@endlink[0]{}%
\providecommand \url  [0]{\begingroup\@sanitize@url \@url }%
\providecommand \@url [1]{\endgroup\@href {#1}{\urlprefix }}%
\providecommand \urlprefix  [0]{URL }%
\providecommand \Eprint [0]{\href }%
\providecommand \doibase [0]{http://dx.doi.org/}%
\providecommand \selectlanguage [0]{\@gobble}%
\providecommand \bibinfo  [0]{\@secondoftwo}%
\providecommand \bibfield  [0]{\@secondoftwo}%
\providecommand \translation [1]{[#1]}%
\providecommand \BibitemOpen [0]{}%
\providecommand \bibitemStop [0]{}%
\providecommand \bibitemNoStop [0]{.\EOS\space}%
\providecommand \EOS [0]{\spacefactor3000\relax}%
\providecommand \BibitemShut  [1]{\csname bibitem#1\endcsname}%
\let\auto@bib@innerbib\@empty
\bibitem [{\citenamefont {Hawking}(2005)}]{Hawking2005PRD}%
  \BibitemOpen
  \bibfield  {author} {\bibinfo {author} {\bibfnamefont {S.~W.}\ \bibnamefont
  {Hawking}},\ }\bibfield  {title} {\enquote {\bibinfo {title} {Information
  loss in black holes},}\ }\href {\doibase 10.1103/PhysRevD.72.084013}
  {\bibfield  {journal} {\bibinfo  {journal} {Phys. Rev. D}\ }\textbf {\bibinfo
  {volume} {72}},\ \bibinfo {pages} {084013} (\bibinfo {year}
  {2005})}\BibitemShut {NoStop}%
\bibitem [{\citenamefont {Nicolini}\ \emph {et~al.}(2006)\citenamefont
  {Nicolini}, \citenamefont {Smailagic},\ and\ \citenamefont
  {Spallucci}}]{Piero2006PLB}%
  \BibitemOpen
  \bibfield  {author} {\bibinfo {author} {\bibfnamefont {P.}~\bibnamefont
  {Nicolini}}, \bibinfo {author} {\bibfnamefont {A.}~\bibnamefont {Smailagic}},
  \ and\ \bibinfo {author} {\bibfnamefont {E.}~\bibnamefont {Spallucci}},\
  }\bibfield  {title} {\enquote {\bibinfo {title} {Noncommutative geometry
  inspired schwarzschild black hole},}\ }\href {\doibase
  https://doi.org/10.1016/j.physletb.2005.11.004} {\bibfield  {journal}
  {\bibinfo  {journal} {Phys. Lett. B}\ }\textbf {\bibinfo {volume} {632}},\
  \bibinfo {pages} {547--551} (\bibinfo {year} {2006})}\BibitemShut {NoStop}%
\bibitem [{\citenamefont {Nicolini}(2009)}]{Nicolini2009IJMPA}%
  \BibitemOpen
  \bibfield  {author} {\bibinfo {author} {\bibfnamefont {P.}~\bibnamefont
  {Nicolini}},\ }\bibfield  {title} {\enquote {\bibinfo {title}
  {{Noncommutative Black Holes, The Final Appeal To Quantum Gravity: A
  Review}},}\ }\href {\doibase 10.1142/S0217751X09043353} {\bibfield  {journal}
  {\bibinfo  {journal} {Int. J. Mod. Phys. A}\ }\textbf {\bibinfo {volume}
  {24}},\ \bibinfo {pages} {1229--1308} (\bibinfo {year} {2009})}\BibitemShut
  {NoStop}%
\bibitem [{\citenamefont {Ghosh}(2018)}]{Ghosh2018CQG}%
  \BibitemOpen
  \bibfield  {author} {\bibinfo {author} {\bibfnamefont {S.~G.}\ \bibnamefont
  {Ghosh}},\ }\bibfield  {title} {\enquote {\bibinfo {title} {{Noncommutative
  geometry inspired Einstein{\textendash}Gauss{\textendash}Bonnet black
  holes}},}\ }\href {\doibase 10.1088/1361-6382/aaaead} {\bibfield  {journal}
  {\bibinfo  {journal} {Class. Quant. Grav.}\ }\textbf {\bibinfo {volume}
  {35}},\ \bibinfo {pages} {085008} (\bibinfo {year} {2018})}\BibitemShut
  {NoStop}%
\bibitem[{\citenamefont{Ara{\'u}jo Filho et al.}(2025)}]{Filho2025PDU}%
  \BibitemOpen
  \bibfield{author} {\bibinfo{author}{\bibfnamefont{A.~A.}\ \bibnamefont{Ara{\'u}jo Filho}},\ \bibinfo{author}{\bibfnamefont{J.~R.}\ \bibnamefont{Nascimento}},\ \bibinfo{author}{\bibfnamefont{A.~Y.}\ \bibnamefont{Petrov}},\ \bibinfo{author}{\bibfnamefont{P.~J.}\ \bibnamefont{Porf{\'\i}rio}},\ \bibinfo{author}{\bibfnamefont{A.}\ \bibnamefont{{\"O}vg{\"u}n}},\ }\bibfield{title} {\enquote{\bibinfo{title}{Properties of an axisymmetric Lorentzian non-commutative black hole}},}\ \href{https://doi.org/10.1016/j.dark.2024.101796}{\bibfield{journal}{\bibinfo{journal}{Phys. Dark Univ.}\ }\textbf{\bibinfo{volume}{47}},\ \bibinfo{pages}{101796} (\bibinfo{year}{2025})}\BibitemShut{NoStop}%
\bibitem[{\citenamefont{Ara{\'u}jo Filho}(2025)}]{Filho2025AP}%
  \BibitemOpen
  \bibfield{author} {\bibinfo{author}{\bibfnamefont{A.~A.}\ \bibnamefont{Ara{\'u}jo Filho}},\ }\bibfield{title} {\enquote{\bibinfo{title}{Particle production induced by a Lorentzian non-commutative spacetime}},}\ \href{https://doi.org/10.1016/j.aop.2025.170167}{\bibfield{journal}{\bibinfo{journal}{Annals Phys.}\ }\textbf{\bibinfo{volume}{481}},\ \bibinfo{pages}{170167} (\bibinfo{year}{2025})}\BibitemShut{NoStop}%
\bibitem[{\citenamefont{Ara{\'u}jo Filho et al.}(2025)}]{Filho2025JCAPKBBH}%
  \BibitemOpen
  \bibfield{author} {\bibinfo{author}{\bibfnamefont{A.~A.}\ \bibnamefont{Ara{\'u}jo Filho}},\ \bibinfo{author}{\bibfnamefont{N.}\ \bibnamefont{Heidari}},\ 
  \bibinfo{author}{\bibfnamefont{Iarley P.}\ \bibnamefont{Lobo}},\ }\bibfield{title} {\enquote{\bibinfo{title}{A non-commutative Kalb-Ramond black hole}},}\ \href{https://doi.org/10.1088/1475-7516/2025/09/076}{\bibfield{journal}{\bibinfo{journal}{JCAP}\ }\textbf{\bibinfo{volume}{09}},\ \bibinfo{pages}{076} (\bibinfo{year}{2025})}\BibitemShut{NoStop}%
\bibitem[{\citenamefont{Ara{\'u}jo Filho et al.}(2025)}]{Filho2025JCAPneutrinos}%
  \BibitemOpen
  \bibfield{author} {\bibinfo{author}{\bibfnamefont{A.~A.}\ \bibnamefont{Ara{\'u}jo Filho}},\ \bibinfo{author}{\bibfnamefont{N.}\ \bibnamefont{Heidari}},\ \bibinfo{author}{\bibfnamefont{Ali}\ \bibnamefont{{\"O}vg{\"u}n}},\ }\bibfield{title} {\enquote{\bibinfo{title}{Geodesics, accretion disk, gravitational lensing, time delay, and effects on neutrinos induced by a non-commutative black hole}},}\ \href{https://doi.org/10.1088/1475-7516/2025/06/062}{\bibfield{journal}{\bibinfo{journal}{JCAP}\ }\textbf{\bibinfo{volume}{06}},\ \bibinfo{pages}{062} (\bibinfo{year}{2025})}\BibitemShut{NoStop}%
\bibitem [{\citenamefont {Rizzo}(2006)}]{Rizzo2006JHEP}%
  \BibitemOpen
  \bibfield  {author} {\bibinfo {author} {\bibfnamefont {T.~G.}\ \bibnamefont
  {Rizzo}},\ }\bibfield  {title} {\enquote {\bibinfo {title} {Noncommutative
  inspired black holes in extra dimensions},}\ }\href {\doibase
  10.1088/1126-6708/2006/09/021} {\bibfield  {journal} {\bibinfo  {journal}
  {JHEP}\ }\textbf {\bibinfo {volume} {2006}},\ \bibinfo {pages} {021}
  (\bibinfo {year} {2006})}\BibitemShut {NoStop}%
\bibitem [{\citenamefont {Ansoldi}\ \emph {et~al.}(2007)\citenamefont
  {Ansoldi}, \citenamefont {Nicolini}, \citenamefont {Smailagic},\ and\
  \citenamefont {Spallucci}}]{Stefano2007PLB}%
  \BibitemOpen
  \bibfield  {author} {\bibinfo {author} {\bibfnamefont {S.}~\bibnamefont
  {Ansoldi}}, \bibinfo {author} {\bibfnamefont {P.}~\bibnamefont {Nicolini}},
  \bibinfo {author} {\bibfnamefont {A.}~\bibnamefont {Smailagic}}, \ and\
  \bibinfo {author} {\bibfnamefont {E.}~\bibnamefont {Spallucci}},\ }\bibfield
  {title} {\enquote {\bibinfo {title} {Non-commutative geometry inspired
  charged black holes},}\ }\href {\doibase https://doi.org/10.1016/j.physletb.2006.12.020} {\bibfield  {journal}
  {\bibinfo  {journal} {Phys. Lett. B}\ }\textbf {\bibinfo {volume} {645}},\
  \bibinfo {pages} {261--266} (\bibinfo {year} {2007})}\BibitemShut {NoStop}%
\bibitem [{\citenamefont {Spallucci}\ \emph {et~al.}(2009)\citenamefont
  {Spallucci}, \citenamefont {Smailagic},\ and\ \citenamefont
  {Nicolini}}]{Euro2009PLB}%
  \BibitemOpen
  \bibfield  {author} {\bibinfo {author} {\bibfnamefont {E.}~\bibnamefont
  {Spallucci}}, \bibinfo {author} {\bibfnamefont {A.}~\bibnamefont
  {Smailagic}}, \ and\ \bibinfo {author} {\bibfnamefont {P.}~\bibnamefont
  {Nicolini}},\ }\bibfield  {title} {\enquote {\bibinfo {title}
  {Non-commutative geometry inspired higher-dimensional charged black holes},}\
  }\href {\doibase https://doi.org/10.1016/j.physletb.2008.11.030} {\bibfield
  {journal} {\bibinfo  {journal} {Phys. Lett. B}\ }\textbf {\bibinfo {volume}
  {670}},\ \bibinfo {pages} {449--454} (\bibinfo {year} {2009})}\BibitemShut
  {NoStop}%
\bibitem [{\citenamefont {Modesto}\ and\ \citenamefont
  {Nicolini}(2010)}]{Modesto2010PRD}%
  \BibitemOpen
  \bibfield  {author} {\bibinfo {author} {\bibfnamefont {L.}~\bibnamefont
  {Modesto}}\ and\ \bibinfo {author} {\bibfnamefont {P.}~\bibnamefont
  {Nicolini}},\ }\bibfield  {title} {\enquote {\bibinfo {title} {Charged
  rotating noncommutative black holes},}\ }\href {\doibase
  10.1103/PhysRevD.82.104035} {\bibfield  {journal} {\bibinfo  {journal} {Phys.
  Rev. D}\ }\textbf {\bibinfo {volume} {82}},\ \bibinfo {pages} {104035}
  (\bibinfo {year} {2010})}\BibitemShut {NoStop}%
\bibitem [{\citenamefont {Cox}\ and\ \citenamefont
  {Gingrich}(2023)}]{Cox2023CQG}%
  \BibitemOpen
  \bibfield  {author} {\bibinfo {author} {\bibfnamefont {Z.}~\bibnamefont
  {Cox}}\ and\ \bibinfo {author} {\bibfnamefont {D.~M.}\ \bibnamefont
  {Gingrich}},\ }\bibfield  {title} {\enquote {\bibinfo {title} {Greybody
  factors for higher-dimensional non-commutative geometry inspired black
  holes},}\ }\href {\doibase 10.1088/1361-6382/aceb46} {\bibfield  {journal}
  {\bibinfo  {journal} {Class. Quant. Grav.}\ }\textbf {\bibinfo {volume}
  {40}},\ \bibinfo {pages} {175013} (\bibinfo {year} {2023})}\BibitemShut
  {NoStop}%
\bibitem [{\citenamefont {Ma}\ \emph {et~al.}(2024)\citenamefont {Ma},
  \citenamefont {Wang}, \citenamefont {Ma}, \citenamefont {Zhang},
  \citenamefont {Deng},\ and\ \citenamefont {Hu}}]{Ma2024EPJP}%
  \BibitemOpen
  \bibfield  {author} {\bibinfo {author} {\bibfnamefont {S.-J.}\ \bibnamefont
  {Ma}}, \bibinfo {author} {\bibfnamefont {R.-B.}\ \bibnamefont {Wang}},
  \bibinfo {author} {\bibfnamefont {T.-C.}\ \bibnamefont {Ma}}, \bibinfo
  {author} {\bibfnamefont {H.-X.}\ \bibnamefont {Zhang}}, \bibinfo {author}
  {\bibfnamefont {J.-B.}\ \bibnamefont {Deng}}, \ and\ \bibinfo {author}
  {\bibfnamefont {X.-R.}\ \bibnamefont {Hu}},\ }\bibfield  {title} {\enquote
  {\bibinfo {title} {Quasinormal modes and greybody factor of charged black
  hole in non-commutative geometry},}\ }\href{https://doi.org/10.1140/epjp/s13360-025-06489-5} {\bibfield  {journal}
  {\bibinfo  {journal} {Eur. Phys. J. Plus}\ }\textbf {\bibinfo {volume}
  {140}},\ \bibinfo {pages} {647} (\bibinfo {year} {2024})}\BibitemShut
  {NoStop}%
\bibitem [{\citenamefont {Ding}\ \emph {et~al.}(2011)\citenamefont {Ding},
  \citenamefont {Kang}, \citenamefont {Chen}, \citenamefont {Chen},\ and\
  \citenamefont {Jing}}]{Ding2010PRD}%
  \BibitemOpen
  \bibfield  {author} {\bibinfo {author} {\bibfnamefont {C.}~\bibnamefont
  {Ding}}, \bibinfo {author} {\bibfnamefont {S.}~\bibnamefont {Kang}}, \bibinfo
  {author} {\bibfnamefont {C.-Y.}\ \bibnamefont {Chen}}, \bibinfo {author}
  {\bibfnamefont {S.}~\bibnamefont {Chen}}, \ and\ \bibinfo {author}
  {\bibfnamefont {J.}~\bibnamefont {Jing}},\ }\bibfield  {title} {\enquote
  {\bibinfo {title} {{Strong gravitational lensing in a noncommutative
  black-hole spacetime}},}\ }\href {\doibase 10.1103/PhysRevD.83.084005}
  {\bibfield  {journal} {\bibinfo  {journal} {Phys. Rev. D}\ }\textbf {\bibinfo
  {volume} {83}},\ \bibinfo {pages} {084005} (\bibinfo {year}
  {2011})}\BibitemShut {NoStop}%
\bibitem [{\citenamefont {Ding}\ and\ \citenamefont
  {Jing}(2011)}]{Ding2011JHEP}%
  \BibitemOpen
  \bibfield  {author} {\bibinfo {author} {\bibfnamefont {C.}~\bibnamefont
  {Ding}}\ and\ \bibinfo {author} {\bibfnamefont {J.}~\bibnamefont {Jing}},\
  }\bibfield  {title} {\enquote {\bibinfo {title} {{Probing spacetime
  noncommutative constant via charged astrophysical black hole lensing}},}\
  }\href {\doibase 10.1007/JHEP10(2011)052} {\bibfield  {journal} {\bibinfo
  {journal} {JHEP}\ }\textbf {\bibinfo {volume} {10}},\ \bibinfo {pages} {052}
  (\bibinfo {year} {2011})}\BibitemShut {NoStop}%
\bibitem [{\citenamefont {Wei}\ \emph {et~al.}(2015)\citenamefont {Wei},
  \citenamefont {Cheng}, \citenamefont {Zhong},\ and\ \citenamefont
  {Zhou}}]{Wei2015JCAP}%
  \BibitemOpen
  \bibfield  {author} {\bibinfo {author} {\bibfnamefont {S.-W.}\ \bibnamefont
  {Wei}}, \bibinfo {author} {\bibfnamefont {P.}~\bibnamefont {Cheng}}, \bibinfo
  {author} {\bibfnamefont {Y.}~\bibnamefont {Zhong}}, \ and\ \bibinfo {author}
  {\bibfnamefont {X.-N.}\ \bibnamefont {Zhou}},\ }\bibfield  {title} {\enquote
  {\bibinfo {title} {{Shadow of noncommutative geometry inspired black
  hole}},}\ }\href {\doibase 10.1088/1475-7516/2015/08/004} {\bibfield
  {journal} {\bibinfo  {journal} {JCAP}\ }\textbf {\bibinfo {volume} {08}},\
  \bibinfo {pages} {004} (\bibinfo {year} {2015})}\BibitemShut {NoStop}%
\bibitem [{\citenamefont {Sharif}\ and\ \citenamefont
  {Iftikhar}(2016)}]{Sharif2016zEPJC}%
  \BibitemOpen
  \bibfield  {author} {\bibinfo {author} {\bibfnamefont {M.}~\bibnamefont
  {Sharif}}\ and\ \bibinfo {author} {\bibfnamefont {S.}~\bibnamefont
  {Iftikhar}},\ }\bibfield  {title} {\enquote {\bibinfo {title} {{Shadow of a
  Charged Rotating Non-Commutative Black Hole}},}\ }\href {\doibase
  10.1140/epjc/s10052-016-4472-3} {\bibfield  {journal} {\bibinfo  {journal}
  {Eur. Phys. J. C}\ }\textbf {\bibinfo {volume} {76}},\ \bibinfo {pages} {630}
  (\bibinfo {year} {2016})}\BibitemShut {NoStop}%
\bibitem [{\citenamefont {Batic}\ \emph {et~al.}(2019)\citenamefont {Batic},
  \citenamefont {Kelkar}, \citenamefont {Nowakowski},\ and\ \citenamefont
  {Redway}}]{Batic2019EPJP}%
  \BibitemOpen
  \bibfield  {author} {\bibinfo {author} {\bibfnamefont {D.}~\bibnamefont
  {Batic}}, \bibinfo {author} {\bibfnamefont {N.~G.}\ \bibnamefont {Kelkar}},
  \bibinfo {author} {\bibfnamefont {M.}~\bibnamefont {Nowakowski}}, \ and\
  \bibinfo {author} {\bibfnamefont {K.}~\bibnamefont {Redway}},\ }\bibfield
  {title} {\enquote {\bibinfo {title} {{Perturbing microscopic black holes
  inspired by noncommutativity}},}\ }\href {\doibase
  10.1140/epjc/s10052-019-7084-x} {\bibfield  {journal} {\bibinfo  {journal}
  {Eur. Phys. J. C}\ }\textbf {\bibinfo {volume} {79}},\ \bibinfo {pages} {581}
  (\bibinfo {year} {2019})}\BibitemShut {NoStop}%
\bibitem [{\citenamefont {{\"O}vg{\"u}n}\ \emph {et~al.}(2020)\citenamefont
  {{\"O}vg{\"u}n}, \citenamefont {Sakall{\i}}, \citenamefont {Saavedra},\ and\
  \citenamefont {Leiva}}]{Ovgun2020MPLA}%
  \BibitemOpen
  \bibfield  {author} {\bibinfo {author} {\bibfnamefont {A.}~\bibnamefont
  {{\"O}vg{\"u}n}}, \bibinfo {author} {\bibfnamefont {{\.I}.}~\bibnamefont
  {Sakall{\i}}}, \bibinfo {author} {\bibfnamefont {J.}~\bibnamefont
  {Saavedra}}, \ and\ \bibinfo {author} {\bibfnamefont {C.}~\bibnamefont
  {Leiva}},\ }\bibfield  {title} {\enquote {\bibinfo {title} {{Shadow cast of
  noncommutative black holes in Rastall gravity}},}\ }\href {\doibase
  10.1142/S0217732320501631} {\bibfield  {journal} {\bibinfo  {journal} {Mod.
  Phys. Lett. A}\ }\textbf {\bibinfo {volume} {35}},\ \bibinfo {pages}
  {2050163} (\bibinfo {year} {2020})}\BibitemShut {NoStop}%
\bibitem [{\citenamefont {Filho}\ \emph {et~al.}(2024)\citenamefont {Filho},
  \citenamefont {Nascimento}, \citenamefont {Petrov}, \citenamefont
  {Porf{\'\i}rio},\ and\ \citenamefont {{\"O}vg{\"u}n}}]{Filho2024PDU}%
  \BibitemOpen
  \bibfield  {author} {\bibinfo {author} {\bibfnamefont {A.~A.~A.}\
  \bibnamefont {Filho}}, \bibinfo {author} {\bibfnamefont {J.~R.}\ \bibnamefont
  {Nascimento}}, \bibinfo {author} {\bibfnamefont {A.~Y.}\ \bibnamefont
  {Petrov}}, \bibinfo {author} {\bibfnamefont {P.~J.}\ \bibnamefont
  {Porf{\'\i}rio}}, \ and\ \bibinfo {author} {\bibfnamefont {A.}~\bibnamefont
  {{\"O}vg{\"u}n}},\ }\bibfield  {title} {\enquote {\bibinfo {title} {{Effects
  of non-commutative geometry on black hole properties}},}\ }\href {\doibase
  10.1016/j.dark.2024.101630} {\bibfield  {journal} {\bibinfo  {journal} {Phys.
  Dark Univ.}\ }\textbf {\bibinfo {volume} {46}},\ \bibinfo {pages} {101630}
  (\bibinfo {year} {2024})}\BibitemShut {NoStop}%
\bibitem [{\citenamefont {Batic}\ and\ \citenamefont
  {Dutykh}(2024)}]{Batic2024EPJC}%
  \BibitemOpen
  \bibfield  {author} {\bibinfo {author} {\bibfnamefont {D.}~\bibnamefont
  {Batic}}\ and\ \bibinfo {author} {\bibfnamefont {D.}~\bibnamefont {Dutykh}},\
  }\bibfield  {title} {\enquote {\bibinfo {title} {Quasinormal modes in
  noncommutative schwarzschild black holes: a spectral analysis},}\ }\href
  {\doibase 10.1140/epjc/s10052-024-12981-6} {\bibfield  {journal} {\bibinfo
  {journal} {Eur. Phys. J. C}\ }\textbf {\bibinfo {volume} {84}},\ \bibinfo
  {pages} {622} (\bibinfo {year} {2024})}\BibitemShut {NoStop}%
\bibitem[{\citenamefont{Heidari et al.}(2025)}]{Heidari2025JCAP}%
  \BibitemOpen
  \bibfield{author} {\bibinfo{author}{\bibfnamefont{N.}\ \bibnamefont{Heidari}},\ \bibinfo{author}{\bibfnamefont{A.~A.}\ \bibnamefont{Ara{\'u}jo Filho}},\ \bibinfo{author}{\bibfnamefont{Iarley P.}\ \bibnamefont{Lobo}},\ }\bibfield{title} {\enquote{\bibinfo{title}{Non-commutativity in Hayward spacetime}},}\ \href{https://doi.org/10.1088/1475-7516/2025/09/051}{\bibfield{journal}{\bibinfo{journal}{JCAP}\ }\textbf{\bibinfo{volume}{09}},\ \bibinfo{pages}{051} (\bibinfo{year}{2025})}\BibitemShut{NoStop}%
\bibitem [{\citenamefont {Vishveshwara}(1970)}]{Vishveshwara1970Scattering}%
  \BibitemOpen
  \bibfield  {author} {\bibinfo {author} {\bibfnamefont {C.~V.}\ \bibnamefont
  {Vishveshwara}},\ }\bibfield  {title} {\enquote {\bibinfo {title} {Scattering
  of gravitational radiation by a schwarzschild black-hole},}\ }\href {\doibase
  10.1038/227936a0} {\bibfield  {journal} {\bibinfo  {journal} {Nature}\
  }\textbf {\bibinfo {volume} {227}},\ \bibinfo {pages} {936--938} (\bibinfo
  {year} {1970})}\BibitemShut {NoStop}%
\bibitem [{\citenamefont {Berti}\ \emph
  {et~al.}(2007{\natexlab{a}})\citenamefont {Berti}, \citenamefont {Cardoso},
  \citenamefont {Gonzalez},\ and\ \citenamefont {Sperhake}}]{Berti2007PRD1}%
  \BibitemOpen
  \bibfield  {author} {\bibinfo {author} {\bibfnamefont {E.}~\bibnamefont
  {Berti}}, \bibinfo {author} {\bibfnamefont {V.}~\bibnamefont {Cardoso}},
  \bibinfo {author} {\bibfnamefont {J.~A.}\ \bibnamefont {Gonzalez}}, \ and\
  \bibinfo {author} {\bibfnamefont {U.}~\bibnamefont {Sperhake}},\ }\bibfield
  {title} {\enquote {\bibinfo {title} {{Mining information from binary black
  hole mergers: A Comparison of estimation methods for complex exponentials in
  noise}},}\ }\href {\doibase 10.1103/PhysRevD.75.124017} {\bibfield  {journal}
  {\bibinfo  {journal} {Phys. Rev. D}\ }\textbf {\bibinfo {volume} {75}},\
  \bibinfo {pages} {124017} (\bibinfo {year} {2007}{\natexlab{a}})}\BibitemShut
  {NoStop}%
\bibitem [{\citenamefont {Echeverria}(1989)}]{Fernando1989PRD}%
  \BibitemOpen
  \bibfield  {author} {\bibinfo {author} {\bibfnamefont {F.}~\bibnamefont
  {Echeverria}},\ }\bibfield  {title} {\enquote {\bibinfo {title}
  {Gravitational-wave measurements of the mass and angular momentum of a black
  hole},}\ }\href {\doibase 10.1103/PhysRevD.40.3194} {\bibfield  {journal}
  {\bibinfo  {journal} {Phys. Rev. D}\ }\textbf {\bibinfo {volume} {40}},\
  \bibinfo {pages} {3194--3203} (\bibinfo {year} {1989})}\BibitemShut {NoStop}%
\bibitem [{\citenamefont {Berti}\ \emph {et~al.}(2006)\citenamefont {Berti},
  \citenamefont {Cardoso},\ and\ \citenamefont {Will}}]{Berti2006PRD}%
  \BibitemOpen
  \bibfield  {author} {\bibinfo {author} {\bibfnamefont {E.}~\bibnamefont
  {Berti}}, \bibinfo {author} {\bibfnamefont {V.}~\bibnamefont {Cardoso}}, \
  and\ \bibinfo {author} {\bibfnamefont {C.~M.}\ \bibnamefont {Will}},\
  }\bibfield  {title} {\enquote {\bibinfo {title} {{On gravitational-wave
  spectroscopy of massive black holes with the space interferometer LISA}},}\
  }\href {\doibase 10.1103/PhysRevD.73.064030} {\bibfield  {journal} {\bibinfo
  {journal} {Phys. Rev. D}\ }\textbf {\bibinfo {volume} {73}},\ \bibinfo
  {pages} {064030} (\bibinfo {year} {2006})}\BibitemShut {NoStop}%
\bibitem [{\citenamefont {Berti}\ \emph
  {et~al.}(2007{\natexlab{b}})\citenamefont {Berti}, \citenamefont {Cardoso},
  \citenamefont {Cardoso},\ and\ \citenamefont {Cavaglia}}]{Berti2007PRD}%
  \BibitemOpen
  \bibfield  {author} {\bibinfo {author} {\bibfnamefont {E.}~\bibnamefont
  {Berti}}, \bibinfo {author} {\bibfnamefont {J.}~\bibnamefont {Cardoso}},
  \bibinfo {author} {\bibfnamefont {V.}~\bibnamefont {Cardoso}}, \ and\
  \bibinfo {author} {\bibfnamefont {M.}~\bibnamefont {Cavaglia}},\ }\bibfield
  {title} {\enquote {\bibinfo {title} {{Matched-filtering and parameter
  estimation of ringdown waveforms}},}\ }\href {\doibase
  10.1103/PhysRevD.76.104044} {\bibfield  {journal} {\bibinfo  {journal} {Phys.
  Rev. D}\ }\textbf {\bibinfo {volume} {76}},\ \bibinfo {pages} {104044}
  (\bibinfo {year} {2007}{\natexlab{b}})}\BibitemShut {NoStop}%
\bibitem [{\citenamefont {Huang}\ \emph {et~al.}(2022)\citenamefont {Huang},
  \citenamefont {Ou}, \citenamefont {Lai},\ and\ \citenamefont
  {Lu}}]{Huang:2021qwe}%
  \BibitemOpen
  \bibfield  {author} {\bibinfo {author} {\bibfnamefont {H.}~\bibnamefont
  {Huang}}, \bibinfo {author} {\bibfnamefont {M.-Y.}\ \bibnamefont {Ou}},
  \bibinfo {author} {\bibfnamefont {M.-Y.}\ \bibnamefont {Lai}}, \ and\
  \bibinfo {author} {\bibfnamefont {H.}~\bibnamefont {Lu}},\ }\bibfield
  {title} {\enquote {\bibinfo {title} {{Echoes from classical black holes}},}\
  }\href {\doibase 10.1103/PhysRevD.105.104049} {\bibfield  {journal} {\bibinfo
   {journal} {Phys. Rev. D}\ }\textbf {\bibinfo {volume} {105}},\ \bibinfo
  {pages} {104049} (\bibinfo {year} {2022})}\BibitemShut {NoStop}%
\bibitem [{\citenamefont {Press}(1971)}]{Press1971AJL}%
  \BibitemOpen
  \bibfield  {author} {\bibinfo {author} {\bibfnamefont {W.~H.}\ \bibnamefont
  {Press}},\ }\bibfield  {title} {\enquote {\bibinfo {title} {{Long Wave Trains
  of Gravitational Waves from a Vibrating Black Hole}},}\ }\href {\doibase
  10.1086/180849} {\bibfield  {journal} {\bibinfo  {journal} {Astrophys. J.
  Lett.}\ }\textbf {\bibinfo {volume} {170}},\ \bibinfo {pages} {L105--L108}
  (\bibinfo {year} {1971})}\BibitemShut {NoStop}%
\bibitem [{\citenamefont {Schutz}\ and\ \citenamefont
  {Will}(1985)}]{Schutz1985AJL}%
  \BibitemOpen
  \bibfield  {author} {\bibinfo {author} {\bibfnamefont {B.~F.}\ \bibnamefont
  {Schutz}}\ and\ \bibinfo {author} {\bibfnamefont {C.~M.}\ \bibnamefont
  {Will}},\ }\bibfield  {title} {\enquote {\bibinfo {title} {{Black hole normal
  modes: A semianalytic approach}},}\ }\href {\doibase 10.1086/184453}
  {\bibfield  {journal} {\bibinfo  {journal} {Astrophys. J. Lett.}\ }\textbf
  {\bibinfo {volume} {291}},\ \bibinfo {pages} {L33--L36} (\bibinfo {year}
  {1985})}\BibitemShut {NoStop}%
  \bibitem [{\citenamefont {Leaver}(1985)}]{Leaver1985AMPS}%
  \BibitemOpen
  \bibfield  {author} {\bibinfo {author} {\bibfnamefont {E.~W.}\ \bibnamefont
  {Leaver}},\ }\bibfield  {title} {\enquote {\bibinfo {title} {An analytic
  representation for the quasi-normal modes of kerr black holes},}\ }\href
  {\doibase 10.1098/rspa.1985.0119} {\bibfield  {journal} {\bibinfo  {journal}
  {Proc. R. Soc. A: Math. Phys. Eng. Sci.}\ }\textbf {\bibinfo {volume}
  {402}},\ \bibinfo {pages} {285--298} (\bibinfo {year} {1985})}\BibitemShut
  {NoStop}%
\bibitem [{\citenamefont {Iyer}\ and\ \citenamefont
  {Will}(1987)}]{Iyer1987PRDI}%
  \BibitemOpen
  \bibfield  {author} {\bibinfo {author} {\bibfnamefont {S.}~\bibnamefont
  {Iyer}}\ and\ \bibinfo {author} {\bibfnamefont {C.~M.}\ \bibnamefont
  {Will}},\ }\bibfield  {title} {\enquote {\bibinfo {title} {Black-hole normal
  modes: A wkb approach. i. foundations and application of a higher-order wkb
  analysis of potential-barrier scattering},}\ }\href {\doibase
  10.1103/PhysRevD.35.3621} {\bibfield  {journal} {\bibinfo  {journal} {Phys.
  Rev. D}\ }\textbf {\bibinfo {volume} {35}},\ \bibinfo {pages} {3621--3631}
  (\bibinfo {year} {1987})}\BibitemShut {NoStop}%
\bibitem [{\citenamefont {Iyer}(1987)}]{Iyer1987PRDII}%
  \BibitemOpen
  \bibfield  {author} {\bibinfo {author} {\bibfnamefont {S.}~\bibnamefont
  {Iyer}},\ }\bibfield  {title} {\enquote {\bibinfo {title} {Black-hole normal
  modes: A wkb approach. ii. schwarzschild black holes},}\ }\href {\doibase
  10.1103/PhysRevD.35.3632} {\bibfield  {journal} {\bibinfo  {journal} {Phys.
  Rev. D}\ }\textbf {\bibinfo {volume} {35}},\ \bibinfo {pages} {3632--3636}
  (\bibinfo {year} {1987})}\BibitemShut {NoStop}%
\bibitem [{\citenamefont {Nollert}\ and\ \citenamefont
  {Schmidt}(1992)}]{Nollert1992PRD}%
  \BibitemOpen
  \bibfield  {author} {\bibinfo {author} {\bibfnamefont {H.-P.}\ \bibnamefont
  {Nollert}}\ and\ \bibinfo {author} {\bibfnamefont {B.~G.}\ \bibnamefont
  {Schmidt}},\ }\bibfield  {title} {\enquote {\bibinfo {title} {Quasinormal
  modes of schwarzschild black holes: Defined and calculated via laplace
  transformation},}\ }\href {\doibase 10.1103/PhysRevD.45.2617} {\bibfield
  {journal} {\bibinfo  {journal} {Phys. Rev. D}\ }\textbf {\bibinfo {volume}
  {45}},\ \bibinfo {pages} {2617--2627} (\bibinfo {year} {1992})}\BibitemShut
  {NoStop}%
\bibitem [{\citenamefont {Fr\"oman}\ \emph {et~al.}(1992)\citenamefont
  {Fr\"oman}, \citenamefont {Fr\"oman}, \citenamefont {Andersson},\ and\
  \citenamefont {H\"okback}}]{Froman1992PRD}%
  \BibitemOpen
  \bibfield  {author} {\bibinfo {author} {\bibfnamefont {N.}~\bibnamefont
  {Fr\"oman}}, \bibinfo {author} {\bibfnamefont {P.~O.}\ \bibnamefont
  {Fr\"oman}}, \bibinfo {author} {\bibfnamefont {N.}~\bibnamefont {Andersson}},
  \ and\ \bibinfo {author} {\bibfnamefont {A.}~\bibnamefont {H\"okback}},\
  }\bibfield  {title} {\enquote {\bibinfo {title} {Black-hole normal modes:
  Phase-integral treatment},}\ }\href {\doibase 10.1103/PhysRevD.45.2609}
  {\bibfield  {journal} {\bibinfo  {journal} {Phys. Rev. D}\ }\textbf {\bibinfo
  {volume} {45}},\ \bibinfo {pages} {2609--2616} (\bibinfo {year}
  {1992})}\BibitemShut {NoStop}%
\bibitem [{\citenamefont {Gundlach}\ \emph {et~al.}(1994)\citenamefont
  {Gundlach}, \citenamefont {Price},\ and\ \citenamefont
  {Pullin}}]{Gundlach1994PRD}%
  \BibitemOpen
  \bibfield  {author} {\bibinfo {author} {\bibfnamefont {C.}~\bibnamefont
  {Gundlach}}, \bibinfo {author} {\bibfnamefont {R.~H.}\ \bibnamefont {Price}},
  \ and\ \bibinfo {author} {\bibfnamefont {J.}~\bibnamefont {Pullin}},\
  }\bibfield  {title} {\enquote {\bibinfo {title} {Late-time behavior of
  stellar collapse and explosions. i. linearized perturbations},}\ }\href
  {\doibase 10.1103/PhysRevD.49.883} {\bibfield  {journal} {\bibinfo  {journal}
  {Phys. Rev. D}\ }\textbf {\bibinfo {volume} {49}},\ \bibinfo {pages}
  {883--889} (\bibinfo {year} {1994})}\BibitemShut {NoStop}%
\bibitem [{\citenamefont {Konoplya}(2003)}]{Konoplya2003PRD}%
  \BibitemOpen
  \bibfield  {author} {\bibinfo {author} {\bibfnamefont {R.~A.}\ \bibnamefont
  {Konoplya}},\ }\bibfield  {title} {\enquote {\bibinfo {title} {{Quasinormal
  behavior of the d-dimensional Schwarzschild black hole and higher order WKB
  approach}},}\ }\href {\doibase 10.1103/PhysRevD.68.024018} {\bibfield
  {journal} {\bibinfo  {journal} {Phys. Rev. D}\ }\textbf {\bibinfo {volume}
  {68}},\ \bibinfo {pages} {024018} (\bibinfo {year} {2003})}\BibitemShut
  {NoStop}%
\bibitem [{\citenamefont {Pani}(2013)}]{Pani2013IJMPA}%
  \BibitemOpen
  \bibfield  {author} {\bibinfo {author} {\bibfnamefont {P.}~\bibnamefont
  {Pani}},\ }\bibfield  {title} {\enquote {\bibinfo {title} {{Advanced Methods
  in Black-Hole Perturbation Theory}},}\ }\href {\doibase
  10.1142/S0217751X13400186} {\bibfield  {journal} {\bibinfo  {journal} {Int.
  J. Mod. Phys. A}\ }\textbf {\bibinfo {volume} {28}},\ \bibinfo {pages}
  {1340018} (\bibinfo {year} {2013})}\BibitemShut {NoStop}%
\bibitem [{\citenamefont {Matyjasek}\ and\ \citenamefont
  {Opala}(2017)}]{Matyjasek2017PRD}%
  \BibitemOpen
  \bibfield  {author} {\bibinfo {author} {\bibfnamefont {J.}~\bibnamefont
  {Matyjasek}}\ and\ \bibinfo {author} {\bibfnamefont {M.}~\bibnamefont
  {Opala}},\ }\bibfield  {title} {\enquote {\bibinfo {title} {{Quasinormal
  modes of black holes. The improved semianalytic approach}},}\ }\href
  {\doibase 10.1103/PhysRevD.96.024011} {\bibfield  {journal} {\bibinfo
  {journal} {Phys. Rev. D}\ }\textbf {\bibinfo {volume} {96}},\ \bibinfo
  {pages} {024011} (\bibinfo {year} {2017})}\BibitemShut {NoStop}%
\bibitem [{\citenamefont {Liang}(2018{\natexlab{a}})}]{Liang2018CPL}%
  \BibitemOpen
  \bibfield  {author} {\bibinfo {author} {\bibfnamefont {J.}~\bibnamefont
  {Liang}},\ }\bibfield  {title} {\enquote {\bibinfo {title} {Quasinormal modes
  of a noncommutative-geometry-inspired schwarzschild black hole},}\ }\href
  {\doibase 10.1088/0256-307X/35/1/010401} {\bibfield  {journal} {\bibinfo
  {journal} {Chin. Phys. Lett.}\ }\textbf {\bibinfo {volume} {35}},\ \bibinfo
  {pages} {010401--010401} (\bibinfo {year} {2018}{\natexlab{a}})}\BibitemShut
  {NoStop}%
\bibitem [{\citenamefont {Liang}(2018{\natexlab{b}})}]{Liang2018CPL2}%
  \BibitemOpen
  \bibfield  {author} {\bibinfo {author} {\bibfnamefont {J.}~\bibnamefont
  {Liang}},\ }\bibfield  {title} {\enquote {\bibinfo {title} {Quasinormal modes
  of a noncommutative-geometry-inspired schwarzschild black hole:
  Gravitational, electromagnetic and massless dirac perturbations*},}\ }\href
  {\doibase 10.1088/0256-307X/35/5/050401} {\bibfield  {journal} {\bibinfo
  {journal} {Chin. Phys. Lett.}\ }\textbf {\bibinfo {volume} {35}},\ \bibinfo
  {pages} {050401} (\bibinfo {year} {2018}{\natexlab{b}})}\BibitemShut
  {NoStop}%
\bibitem [{\citenamefont {Karimabadi}\ \emph {et~al.}(2025)\citenamefont
  {Karimabadi}, \citenamefont {Yekta},\ and\ \citenamefont
  {Alavi}}]{Karimabadi2025arXiv}%
  \BibitemOpen
  \bibfield  {author} {\bibinfo {author} {\bibfnamefont {M.}~\bibnamefont
  {Karimabadi}}, \bibinfo {author} {\bibfnamefont {D.~M.}\ \bibnamefont
  {Yekta}}, \ and\ \bibinfo {author} {\bibfnamefont {S.~A.}\ \bibnamefont
  {Alavi}},\ }\bibfield  {title} {\enquote {\bibinfo {title} {{Effects of
  non-minimal scalar field couplings with curvature tensors on perturbations in
  non-commutative Schwarzschild spacetimes}},}\ }\href@noop {} {\  (\bibinfo
  {year} {2025})},\ \Eprint {http://arxiv.org/abs/2508.13820} {arXiv:2508.13820
  [gr-qc]} \BibitemShut {NoStop}%
\bibitem [{\citenamefont {Fan}\ \emph {et~al.}(2025)\citenamefont {Fan},
  \citenamefont {Wu},\ and\ \citenamefont {Guo}}]{Fan2025arXiv}%
  \BibitemOpen
  \bibfield  {author} {\bibinfo {author} {\bibfnamefont {S.-H.}\ \bibnamefont
  {Fan}}, \bibinfo {author} {\bibfnamefont {C.}~\bibnamefont {Wu}}, \ and\
  \bibinfo {author} {\bibfnamefont {W.-J.}\ \bibnamefont {Guo}},\ }\bibfield
  {title} {\enquote {\bibinfo {title} {{Grey-body Factors and Absorption Cross
  Sections of Non-Commutative Black Holes under Einstein-Coupled Scalar
  Fields}},}\ }\href@noop {} {\  (\bibinfo {year} {2025})},\ \Eprint
  {http://arxiv.org/abs/2511.16012} {arXiv:2511.16012 [gr-qc]} \BibitemShut
  {NoStop}%
\bibitem [{\citenamefont {Sanchez}(1978)}]{Sanchez1978PRD}%
  \BibitemOpen
  \bibfield  {author} {\bibinfo {author} {\bibfnamefont {N.}~\bibnamefont
  {Sanchez}},\ }\bibfield  {title} {\enquote {\bibinfo {title} {Absorption and
  emission spectra of a schwarzschild black hole},}\ }\href {\doibase
  10.1103/PhysRevD.18.1030} {\bibfield  {journal} {\bibinfo  {journal} {Phys.
  Rev. D}\ }\textbf {\bibinfo {volume} {18}},\ \bibinfo {pages} {1030--1036}
  (\bibinfo {year} {1978})}\BibitemShut {NoStop}%
\bibitem [{\citenamefont {Andersson}(1995)}]{Andersson1995PRD}%
  \BibitemOpen
  \bibfield  {author} {\bibinfo {author} {\bibfnamefont {N.}~\bibnamefont
  {Andersson}},\ }\bibfield  {title} {\enquote {\bibinfo {title} {Scattering of
  massless scalar waves by a schwarzschild black hole: A phase-integral
  study},}\ }\href {\doibase 10.1103/PhysRevD.52.1808} {\bibfield  {journal}
  {\bibinfo  {journal} {Phys. Rev. D}\ }\textbf {\bibinfo {volume} {52}},\
  \bibinfo {pages} {1808--1820} (\bibinfo {year} {1995})}\BibitemShut {NoStop}%
\bibitem [{\citenamefont {Kanti}\ and\ \citenamefont
  {March-Russell}(2002)}]{Kanti2002PRD}%
  \BibitemOpen
  \bibfield  {author} {\bibinfo {author} {\bibfnamefont {P.}~\bibnamefont
  {Kanti}}\ and\ \bibinfo {author} {\bibfnamefont {J.}~\bibnamefont
  {March-Russell}},\ }\bibfield  {title} {\enquote {\bibinfo {title}
  {{Calculable corrections to brane black hole decay. 1. The scalar case}},}\
  }\href {\doibase 10.1103/PhysRevD.66.024023} {\bibfield  {journal} {\bibinfo
  {journal} {Phys. Rev. D}\ }\textbf {\bibinfo {volume} {66}},\ \bibinfo
  {pages} {024023} (\bibinfo {year} {2002})}\BibitemShut {NoStop}%
\bibitem [{\citenamefont {Cardoso}\ \emph {et~al.}(2006)\citenamefont
  {Cardoso}, \citenamefont {Cavaglia},\ and\ \citenamefont
  {Gualtieri}}]{Cardoso2006PRL}%
  \BibitemOpen
  \bibfield  {author} {\bibinfo {author} {\bibfnamefont {V.}~\bibnamefont
  {Cardoso}}, \bibinfo {author} {\bibfnamefont {M.}~\bibnamefont {Cavaglia}}, \
  and\ \bibinfo {author} {\bibfnamefont {L.}~\bibnamefont {Gualtieri}},\
  }\bibfield  {title} {\enquote {\bibinfo {title} {{Black Hole Particle
  Emission in Higher-Dimensional Spacetimes}},}\ }\href {\doibase
  10.1103/PhysRevLett.96.071301} {\bibfield  {journal} {\bibinfo  {journal}
  {Phys. Rev. Lett.}\ }\textbf {\bibinfo {volume} {96}},\ \bibinfo {pages}
  {071301} (\bibinfo {year} {2006})}\BibitemShut {NoStop}%
\bibitem [{\citenamefont {Konoplya}\ and\ \citenamefont
  {Zinhailo}(2019)}]{Konoplya2019PRD}%
  \BibitemOpen
  \bibfield  {author} {\bibinfo {author} {\bibfnamefont {R.~A.}\ \bibnamefont
  {Konoplya}}\ and\ \bibinfo {author} {\bibfnamefont {A.~F.}\ \bibnamefont
  {Zinhailo}},\ }\bibfield  {title} {\enquote {\bibinfo {title} {{Hawking
  radiation of non-Schwarzschild black holes in higher derivative gravity: a
  crucial role of grey-body factors}},}\ }\href {\doibase
  10.1103/PhysRevD.99.104060} {\bibfield  {journal} {\bibinfo  {journal} {Phys.
  Rev. D}\ }\textbf {\bibinfo {volume} {99}},\ \bibinfo {pages} {104060}
  (\bibinfo {year} {2019})}\BibitemShut {NoStop}%
\bibitem [{\citenamefont {Mashhoon}(1973)}]{Mashhoon1973PRD}%
  \BibitemOpen
  \bibfield  {author} {\bibinfo {author} {\bibfnamefont {B.}~\bibnamefont
  {Mashhoon}},\ }\bibfield  {title} {\enquote {\bibinfo {title} {Scattering of
  electromagnetic radiation from a black hole},}\ }\href {\doibase
  10.1103/PhysRevD.7.2807} {\bibfield  {journal} {\bibinfo  {journal} {Phys.
  Rev. D}\ }\textbf {\bibinfo {volume} {7}},\ \bibinfo {pages} {2807--2814}
  (\bibinfo {year} {1973})}\BibitemShut {NoStop}%
\bibitem [{\citenamefont {Fabbri}(1975)}]{Fabbri1975PRD}%
  \BibitemOpen
  \bibfield  {author} {\bibinfo {author} {\bibfnamefont {R.}~\bibnamefont
  {Fabbri}},\ }\bibfield  {title} {\enquote {\bibinfo {title} {Scattering and
  absorption of electromagnetic waves by a schwarzschild black hole},}\ }\href
  {\doibase 10.1103/PhysRevD.12.933} {\bibfield  {journal} {\bibinfo  {journal}
  {Phys. Rev. D}\ }\textbf {\bibinfo {volume} {12}},\ \bibinfo {pages}
  {933--942} (\bibinfo {year} {1975})}\BibitemShut {NoStop}%
\bibitem [{\citenamefont {Ould El~Hadj}(2025)}]{Ould2025PRD}%
  \BibitemOpen
  \bibfield  {author} {\bibinfo {author} {\bibfnamefont {M.}~\bibnamefont {Ould
  El~Hadj}},\ }\bibfield  {title} {\enquote {\bibinfo {title} {{Black hole
  absorption cross sections: Spin and Regge poles}},}\ }\href {\doibase
  10.1103/vj91-h7wd} {\bibfield  {journal} {\bibinfo  {journal} {Phys. Rev. D}\
  }\textbf {\bibinfo {volume} {111}},\ \bibinfo {pages} {124041} (\bibinfo
  {year} {2025})}\BibitemShut {NoStop}%
\bibitem [{\citenamefont {Ohashi}\ and\ \citenamefont
  {a.~Sakagami}(2004)}]{Ohashi2004CQG}%
  \BibitemOpen
  \bibfield  {author} {\bibinfo {author} {\bibfnamefont {A.}~\bibnamefont
  {Ohashi}}\ and\ \bibinfo {author} {\bibfnamefont {M.}~\bibnamefont
  {a.~Sakagami}},\ }\bibfield  {title} {\enquote {\bibinfo {title} {{Massive
  quasi-normal mode}},}\ }\href {\doibase 10.1088/0264-9381/21/16/010}
  {\bibfield  {journal} {\bibinfo  {journal} {Class. Quant. Grav.}\ }\textbf
  {\bibinfo {volume} {21}},\ \bibinfo {pages} {3973--3984} (\bibinfo {year}
  {2004})}\BibitemShut {NoStop}%
\bibitem [{\citenamefont {Konoplya}\ and\ \citenamefont
  {Zhidenko}(2005)}]{Konoplya2005PLB}%
  \BibitemOpen
  \bibfield  {author} {\bibinfo {author} {\bibfnamefont {R.~A.}\ \bibnamefont
  {Konoplya}}\ and\ \bibinfo {author} {\bibfnamefont {A.~V.}\ \bibnamefont
  {Zhidenko}},\ }\bibfield  {title} {\enquote {\bibinfo {title} {{Decay of
  massive scalar field in a Schwarzschild background}},}\ }\href {\doibase 10.1016/j.physletb.2005.01.078} {\bibfield  {journal} {\bibinfo  {journal}
  {Phys. Lett. B}\ }\textbf {\bibinfo {volume} {609}},\ \bibinfo {pages}
  {377--384} (\bibinfo {year} {2005})}\BibitemShut {NoStop}%
\bibitem [{\citenamefont {Zhidenko}(2006)}]{Zhidenko2006PRD}%
  \BibitemOpen
  \bibfield  {author} {\bibinfo {author} {\bibfnamefont {A.}~\bibnamefont
  {Zhidenko}},\ }\bibfield  {title} {\enquote {\bibinfo {title} {Massive scalar
  field quasinormal modes of higher dimensional black holes},}\ }\href
  {\doibase 10.1103/PhysRevD.74.064017} {\bibfield  {journal} {\bibinfo
  {journal} {Phys. Rev. D}\ }\textbf {\bibinfo {volume} {74}},\ \bibinfo
  {pages} {064017} (\bibinfo {year} {2006})}\BibitemShut {NoStop}%
\bibitem [{\citenamefont {Konoplya}(2006)}]{Konoplya2006PRD}%
  \BibitemOpen
  \bibfield  {author} {\bibinfo {author} {\bibfnamefont {R.~A.}\ \bibnamefont
  {Konoplya}},\ }\bibfield  {title} {\enquote {\bibinfo {title} {{Massive
  vector field perturbations in the Schwarzschild background: Stability and
  unusual quasinormal spectrum}},}\ }\href {\doibase
  10.1103/PhysRevD.73.024009} {\bibfield  {journal} {\bibinfo  {journal} {Phys.
  Rev. D}\ }\textbf {\bibinfo {volume} {73}},\ \bibinfo {pages} {024009}
  (\bibinfo {year} {2006})}\BibitemShut {NoStop}%
\bibitem [{\citenamefont {Konoplya}\ \emph {et~al.}(2018)\citenamefont
  {Konoplya}, \citenamefont {Stuchl{\'\i}k},\ and\ \citenamefont
  {Zhidenko}}]{Konoplya2018PRD}%
  \BibitemOpen
  \bibfield  {author} {\bibinfo {author} {\bibfnamefont {R.~A.}\ \bibnamefont
  {Konoplya}}, \bibinfo {author} {\bibfnamefont {Z.}~\bibnamefont
  {Stuchl{\'\i}k}}, \ and\ \bibinfo {author} {\bibfnamefont {A.}~\bibnamefont
  {Zhidenko}},\ }\bibfield  {title} {\enquote {\bibinfo {title} {{Axisymmetric
  black holes allowing for separation of variables in the Klein-Gordon and
  Hamilton-Jacobi equations}},}\ }\href {\doibase 10.1103/PhysRevD.97.084044}
  {\bibfield  {journal} {\bibinfo  {journal} {Phys. Rev. D}\ }\textbf {\bibinfo
  {volume} {97}},\ \bibinfo {pages} {084044} (\bibinfo {year}
  {2018})}\BibitemShut {NoStop}%
\bibitem [{\citenamefont {Konoplya}\ and\ \citenamefont
  {Zhidenko}(2018)}]{Konoplya2018PRDlonglived}%
  \BibitemOpen
  \bibfield  {author} {\bibinfo {author} {\bibfnamefont {R.~A.}\ \bibnamefont
  {Konoplya}}\ and\ \bibinfo {author} {\bibfnamefont {A.}~\bibnamefont
  {Zhidenko}},\ }\bibfield  {title} {\enquote {\bibinfo {title} {{Quasinormal
  modes of massive fermions in Kerr spacetime: Long-lived modes and the fine
  structure}},}\ }\href {\doibase 10.1103/PhysRevD.97.084034} {\bibfield
  {journal} {\bibinfo  {journal} {Phys. Rev. D}\ }\textbf {\bibinfo {volume}
  {97}},\ \bibinfo {pages} {084034} (\bibinfo {year} {2018})}\BibitemShut
  {NoStop}%
\bibitem [{\citenamefont {Zhang}\ \emph {et~al.}(2019)\citenamefont {Zhang},
  \citenamefont {Jiang},\ and\ \citenamefont {Zhong}}]{Zhang2019PLB}%
  \BibitemOpen
  \bibfield  {author} {\bibinfo {author} {\bibfnamefont {M.}~\bibnamefont
  {Zhang}}, \bibinfo {author} {\bibfnamefont {J.}~\bibnamefont {Jiang}}, \ and\
  \bibinfo {author} {\bibfnamefont {Z.}~\bibnamefont {Zhong}},\ }\bibfield
  {title} {\enquote {\bibinfo {title} {{The longlived charged massive scalar
  field in the higher-dimensional Reissner{\textendash}Nordstr{\"o}m
  spacetime}},}\ }\href {\doibase 10.1016/j.physletb.2018.10.072} {\bibfield
  {journal} {\bibinfo  {journal} {Phys. Lett. B}\ }\textbf {\bibinfo {volume}
  {789}},\ \bibinfo {pages} {13--18} (\bibinfo {year} {2019})}\BibitemShut
  {NoStop}%
\bibitem [{\citenamefont {Arag{\'o}n}\ \emph {et~al.}(2021)\citenamefont
  {Arag{\'o}n}, \citenamefont {B{\'e}car}, \citenamefont {Gonz{\'a}lez},\ and\
  \citenamefont {V{\'a}squez}}]{Aragon2021PRD}%
  \BibitemOpen
  \bibfield  {author} {\bibinfo {author} {\bibfnamefont {A.}~\bibnamefont
  {Arag{\'o}n}}, \bibinfo {author} {\bibfnamefont {R.}~\bibnamefont
  {B{\'e}car}}, \bibinfo {author} {\bibfnamefont {P.~A.}\ \bibnamefont
  {Gonz{\'a}lez}}, \ and\ \bibinfo {author} {\bibfnamefont {Y.}~\bibnamefont
  {V{\'a}squez}},\ }\bibfield  {title} {\enquote {\bibinfo {title} {{Massive
  Dirac quasinormal modes in Schwarzschild{\textendash}de Sitter black holes:
  Anomalous decay rate and fine structure}},}\ }\href {\doibase
  10.1103/PhysRevD.103.064006} {\bibfield  {journal} {\bibinfo  {journal}
  {Phys. Rev. D}\ }\textbf {\bibinfo {volume} {103}},\ \bibinfo {pages}
  {064006} (\bibinfo {year} {2021})}\BibitemShut {NoStop}%
\bibitem [{\citenamefont {L{\"u}tf{\"u}o{\u{g}}lu}(2026)}]{Lutfuoglu2026arXiv}%
  \BibitemOpen
  \bibfield  {author} {\bibinfo {author} {\bibfnamefont {B.~C.}\ \bibnamefont
  {L{\"u}tf{\"u}o{\u{g}}lu}},\ }\bibfield  {title} {\enquote {\bibinfo {title}
  {{Long-lived quasinormal modes, shadows and particle motion in
  four-dimensional quasi-topological gravity}},}\ }\href@noop {} {\  (\bibinfo
  {year} {2026})},\ \Eprint {http://arxiv.org/abs/2603.10844} {arXiv:2603.10844
  [gr-qc]} \BibitemShut {NoStop}%
\bibitem [{\citenamefont {Smailagic}\ and\ \citenamefont
  {Spallucci}(2003)}]{Smailagic2003JPA}%
  \BibitemOpen
  \bibfield  {author} {\bibinfo {author} {\bibfnamefont {A.}~\bibnamefont
  {Smailagic}}\ and\ \bibinfo {author} {\bibfnamefont {E.}~\bibnamefont
  {Spallucci}},\ }\bibfield  {title} {\enquote {\bibinfo {title} {{Feynman path
  integral on the noncommutative plane}},}\ }\href {\doibase
  10.1088/0305-4470/36/33/101} {\bibfield  {journal} {\bibinfo  {journal} {J.
  Phys. A}\ }\textbf {\bibinfo {volume} {36}},\ \bibinfo {pages} {L467}
  (\bibinfo {year} {2003})}\BibitemShut {NoStop}%
\bibitem [{\citenamefont {Luo}\ \emph {et~al.}(2015)\citenamefont {Luo},
  \citenamefont {Hou}, \citenamefont {Cui}, \citenamefont {Liu},\ and\
  \citenamefont {Zong}}]{Luo:2014iha}%
  \BibitemOpen
  \bibfield  {author} {\bibinfo {author} {\bibfnamefont {C.-B.}\ \bibnamefont
  {Luo}}, \bibinfo {author} {\bibfnamefont {F.-Y.}\ \bibnamefont {Hou}},
  \bibinfo {author} {\bibfnamefont {Z.-F.}\ \bibnamefont {Cui}}, \bibinfo
  {author} {\bibfnamefont {X.-J.}\ \bibnamefont {Liu}}, \ and\ \bibinfo
  {author} {\bibfnamefont {H.-S.}\ \bibnamefont {Zong}},\ }\bibfield  {title}
  {\enquote {\bibinfo {title} {{Noncommutative field with constant background
  fields and neutral fermions}},}\ }\href {\doibase 10.1103/PhysRevD.91.036009}
  {\bibfield  {journal} {\bibinfo  {journal} {Phys. Rev. D}\ }\textbf {\bibinfo
  {volume} {91}},\ \bibinfo {pages} {036009} (\bibinfo {year}
  {2015})}\BibitemShut {NoStop}%
\bibitem [{\citenamefont {Panotopoulos}\ and\ \citenamefont
  {Rinc{\'o}n}(2020)}]{Panotopoulos2020EPJP}%
  \BibitemOpen
  \bibfield  {author} {\bibinfo {author} {\bibfnamefont {G.}~\bibnamefont
  {Panotopoulos}}\ and\ \bibinfo {author} {\bibfnamefont {{\'A}.}~\bibnamefont
  {Rinc{\'o}n}},\ }\bibfield  {title} {\enquote {\bibinfo {title} {{Quasinormal
  modes of five-dimensional black holes in non-commutative geometry}},}\ }\href
  {\doibase 10.1140/epjp/s13360-019-00016-z} {\bibfield  {journal} {\bibinfo
  {journal} {Eur. Phys. J. Plus}\ }\textbf {\bibinfo {volume} {135}},\ \bibinfo
  {pages} {33} (\bibinfo {year} {2020})}\BibitemShut {NoStop}%
\bibitem[{\citenamefont{Das et al.}(2019)}]{Das2019PRD}%
  \BibitemOpen
  \bibfield{author} {\bibinfo{author}{\bibfnamefont{K.}\ \bibnamefont{Das}},\ \bibinfo{author}{\bibfnamefont{S.}\ \bibnamefont{Pramanik}},\ \bibinfo{author}{\bibfnamefont{S.}\ \bibnamefont{Ghosh}},\ }\bibfield{title} {\enquote{\bibinfo{title}{Quasinormal mode spectra for odd parity perturbations in spacetimes with smeared matter sources}},}\ \href{https://doi.org/10.1103/PhysRevD.99.024039}{\bibfield{journal}{\bibinfo{journal}{Phys. Rev. D}\ }\textbf{\bibinfo{volume}{99}},\ \bibinfo{number}{2},\ \bibinfo{pages}{024039} (\bibinfo{year}{2019})}\BibitemShut{NoStop}%
\bibitem [{\citenamefont {Kempf}\ \emph {et~al.}(1995)\citenamefont {Kempf},
  \citenamefont {Mangano},\ and\ \citenamefont {Mann}}]{Kempf1994PRD}%
  \BibitemOpen
  \bibfield  {author} {\bibinfo {author} {\bibfnamefont {A.}~\bibnamefont
  {Kempf}}, \bibinfo {author} {\bibfnamefont {G.}~\bibnamefont {Mangano}}, \
  and\ \bibinfo {author} {\bibfnamefont {R.~B.}\ \bibnamefont {Mann}},\
  }\bibfield  {title} {\enquote {\bibinfo {title} {{Hilbert space
  representation of the minimal length uncertainty relation}},}\ }\href
  {\doibase 10.1103/PhysRevD.52.1108} {\bibfield  {journal} {\bibinfo
  {journal} {Phys. Rev. D}\ }\textbf {\bibinfo {volume} {52}},\ \bibinfo
  {pages} {1108--1118} (\bibinfo {year} {1995})}\BibitemShut {NoStop}%
\bibitem [{\citenamefont {Scardigli}(1999)}]{Scardigli1999PLB}%
  \BibitemOpen
  \bibfield  {author} {\bibinfo {author} {\bibfnamefont {F.}~\bibnamefont
  {Scardigli}},\ }\bibfield  {title} {\enquote {\bibinfo {title} {{Generalized
  uncertainty principle in quantum gravity from micro - black hole Gedanken
  experiment}},}\ }\href {\doibase 10.1016/S0370-2693(99)00167-7} {\bibfield
  {journal} {\bibinfo  {journal} {Phys. Lett. B}\ }\textbf {\bibinfo {volume}
  {452}},\ \bibinfo {pages} {39--44} (\bibinfo {year} {1999})}\BibitemShut
  {NoStop}%
\bibitem [{\citenamefont {Bagrov}\ and\ \citenamefont
  {Obukhov}(1990)}]{Bagrov1990CQG}%
  \BibitemOpen
  \bibfield  {author} {\bibinfo {author} {\bibfnamefont {V.~G.}\ \bibnamefont
  {Bagrov}}\ and\ \bibinfo {author} {\bibfnamefont {V.~V.}\ \bibnamefont
  {Obukhov}},\ }\bibfield  {title} {\enquote {\bibinfo {title} {Separation of
  variables for the klein-gordon equation in special stackel spacetimes},}\
  }\href {\doibase 10.1088/0264-9381/7/1/008} {\bibfield  {journal} {\bibinfo
  {journal} {Class. Quant. Grav.}\ }\textbf {\bibinfo {volume} {7}},\ \bibinfo
  {pages} {19} (\bibinfo {year} {1990})}\BibitemShut {NoStop}%
\bibitem [{\citenamefont {Konoplya}\ and\ \citenamefont
  {Zhidenko}(2011)}]{Konoplya2011RMP}%
  \BibitemOpen
  \bibfield  {author} {\bibinfo {author} {\bibfnamefont {R.~A.}\ \bibnamefont
  {Konoplya}}\ and\ \bibinfo {author} {\bibfnamefont {A.}~\bibnamefont
  {Zhidenko}},\ }\bibfield  {title} {\enquote {\bibinfo {title} {{Quasinormal
  modes of black holes: From astrophysics to string theory}},}\ }\href
  {\doibase 10.1103/RevModPhys.83.793} {\bibfield  {journal} {\bibinfo
  {journal} {Rev. Mod. Phys.}\ }\textbf {\bibinfo {volume} {83}},\ \bibinfo
  {pages} {793--836} (\bibinfo {year} {2011})}\BibitemShut {NoStop}%
\bibitem [{\citenamefont {L\"utf\"uo\u{g}lu}\ \emph {et~al.}(2025)\citenamefont
  {L\"utf\"uo\u{g}lu}, \citenamefont {Saka}, \citenamefont {Shermatov} \emph
  {et~al.}}]{Lutfiuoglu2025EPJC}%
  \BibitemOpen
  \bibfield  {author} {\bibinfo {author} {\bibfnamefont {B.~C.}\ \bibnamefont
  {L\"utf\"uo\u{g}lu}}, \bibinfo {author} {\bibfnamefont {E.~U.}\ \bibnamefont
  {Saka}}, \bibinfo {author} {\bibfnamefont {A.}~\bibnamefont {Shermatov}},
  \emph {et~al.},\ }\bibfield  {title} {\enquote {\bibinfo {title} {Proper-time
  approach in asymptotic safety via black hole quasinormal modes and grey-body
  factors},}\ }\href {\doibase 10.1140/epjc/s10052-025-14950-z} {\bibfield
  {journal} {\bibinfo  {journal} {Eur. Phys. J. C}\ }\textbf {\bibinfo {volume}
  {85}},\ \bibinfo {pages} {1190} (\bibinfo {year} {2025})}\BibitemShut
  {NoStop}%
\bibitem [{\citenamefont {Konoplya}\ \emph {et~al.}(2019)\citenamefont
  {Konoplya}, \citenamefont {Zhidenko},\ and\ \citenamefont
  {Zinhailo}}]{Konoplya2019CQG}%
  \BibitemOpen
  \bibfield  {author} {\bibinfo {author} {\bibfnamefont {R.~A.}\ \bibnamefont
  {Konoplya}}, \bibinfo {author} {\bibfnamefont {A.}~\bibnamefont {Zhidenko}},
  \ and\ \bibinfo {author} {\bibfnamefont {A.~F.}\ \bibnamefont {Zinhailo}},\
  }\bibfield  {title} {\enquote {\bibinfo {title} {{Higher order WKB formula
  for quasinormal modes and grey-body factors: recipes for quick and accurate
  calculations}},}\ }\href {\doibase 10.1088/1361-6382/ab2e25} {\bibfield
  {journal} {\bibinfo  {journal} {Class. Quant. Grav.}\ }\textbf {\bibinfo
  {volume} {36}},\ \bibinfo {pages} {155002} (\bibinfo {year}
  {2019})}\BibitemShut {NoStop}%
\bibitem [{\citenamefont {Tang}\ \emph {et~al.}(2025)\citenamefont {Tang},
  \citenamefont {Ling},\ and\ \citenamefont {Jiang}}]{Chen2025CPC}%
  \BibitemOpen
  \bibfield  {author} {\bibinfo {author} {\bibfnamefont {C.}~\bibnamefont
  {Tang}}, \bibinfo {author} {\bibfnamefont {Y.}~\bibnamefont {Ling}}, \ and\
  \bibinfo {author} {\bibfnamefont {Q.-Q.}\ \bibnamefont {Jiang}},\ }\bibfield
  {title} {\enquote {\bibinfo {title} {Correspondence between grey-body factors
  and quasinormal modes for regular black holes with sub-planckian
  curvature},}\ }\href {\doibase 10.1088/1674-1137/adfa74} {\bibfield
  {journal} {\bibinfo  {journal} {Chin. Phys. C}\ }\textbf {\bibinfo {volume}
  {49}},\ \bibinfo {pages} {125110} (\bibinfo {year} {2025})}\BibitemShut
  {NoStop}%
\bibitem [{\citenamefont {Zhang}\ \emph {et~al.}(2026)\citenamefont {Zhang},
  \citenamefont {Chen}, \citenamefont {Zhang}, \citenamefont {Li},
  \citenamefont {Zhang},\ and\ \citenamefont {Zou}}]{Zhang2026CTP}%
  \BibitemOpen
  \bibfield  {author} {\bibinfo {author} {\bibfnamefont {M.}~\bibnamefont
  {Zhang}}, \bibinfo {author} {\bibfnamefont {G.-X.}\ \bibnamefont {Chen}},
  \bibinfo {author} {\bibfnamefont {L.}~\bibnamefont {Zhang}}, \bibinfo
  {author} {\bibfnamefont {S.-Y.}\ \bibnamefont {Li}}, \bibinfo {author}
  {\bibfnamefont {X.}~\bibnamefont {Zhang}}, \ and\ \bibinfo {author}
  {\bibfnamefont {D.-C.}\ \bibnamefont {Zou}},\ }\bibfield  {title} {\enquote
  {\bibinfo {title} {{Quasinormal modes and greybody factors of magnetically
  charged de Sitter black holes probed by massless external fields in
  Einstein{\textendash}Euler{\textendash}Heisenberg gravity}},}\ }\href
  {\doibase 10.1088/1572-9494/ae42b1} {\bibfield  {journal} {\bibinfo
  {journal} {Commun. Theor. Phys.}\ }\textbf {\bibinfo {volume} {78}},\
  \bibinfo {pages} {055406} (\bibinfo {year} {2026})}\BibitemShut {NoStop}%
\bibitem [{\citenamefont {Arbelaez}(2026)}]{Arbelaez2026arXiv}%
  \BibitemOpen
  \bibfield  {author} {\bibinfo {author} {\bibfnamefont {J.~P.}\ \bibnamefont
  {Arbelaez}},\ }\bibfield  {title} {\enquote {\bibinfo {title} {{Grey-body
  factors of higher dimensional regular black holes in quasi-topological
  theories}},}\ }\href@noop {} {\  (\bibinfo {year} {2026})},\ \Eprint
  {http://arxiv.org/abs/2601.22340} {arXiv:2601.22340 [gr-qc]} \BibitemShut
  {NoStop}%
\bibitem [{\citenamefont {B{\'e}car}\ \emph {et~al.}(2026)\citenamefont
  {B{\'e}car}, \citenamefont {Gonz{\'a}lez}, \citenamefont {Papantonopoulos},\
  and\ \citenamefont {V{\'a}squez}}]{Becar2026arXiv}%
  \BibitemOpen
  \bibfield  {author} {\bibinfo {author} {\bibfnamefont {R.}~\bibnamefont
  {B{\'e}car}}, \bibinfo {author} {\bibfnamefont {P.~A.}\ \bibnamefont
  {Gonz{\'a}lez}}, \bibinfo {author} {\bibfnamefont {E.}~\bibnamefont
  {Papantonopoulos}}, \ and\ \bibinfo {author} {\bibfnamefont {Y.}~\bibnamefont
  {V{\'a}squez}},\ }\bibfield  {title} {\enquote {\bibinfo {title} {{Anomalous
  Decay Rate and Greybody Factors for Regular Black Holes with Scalar Hair}},}\
  }\href@noop {} {\  (\bibinfo {year} {2026})},\ \Eprint
  {http://arxiv.org/abs/2602.16972} {arXiv:2602.16972 [gr-qc]} \BibitemShut
  {NoStop}%
\bibitem [{\citenamefont {Tang}\ \emph {et~al.}(2026)\citenamefont {Tang},
  \citenamefont {Huang},\ and\ \citenamefont {Zhang}}]{Tang2026arXiv}%
  \BibitemOpen
  \bibfield  {author} {\bibinfo {author} {\bibfnamefont {J.}~\bibnamefont
  {Tang}}, \bibinfo {author} {\bibfnamefont {Y.}~\bibnamefont {Huang}}, \ and\
  \bibinfo {author} {\bibfnamefont {H.}~\bibnamefont {Zhang}},\ }\bibfield
  {title} {\enquote {\bibinfo {title} {{Absorption and scattering of massless
  scalar waves by Frolov black holes}},}\ }\href@noop {} {\  (\bibinfo {year}
  {2026})},\ \Eprint {http://arxiv.org/abs/2601.19364} {arXiv:2601.19364
  [gr-qc]} \BibitemShut {NoStop}%
\end{thebibliography}
\end{document}